\documentclass[fleqn,usenatbib]{mnras}

\usepackage{newtxtext,newtxmath}

\usepackage[T1]{fontenc}

\DeclareRobustCommand{\VAN}[3]{#2}
\let\VANthebibliography\thebibliography
\def\thebibliography{\DeclareRobustCommand{\VAN}[3]{##3}\VANthebibliography}

\usepackage{graphicx}	
\usepackage{amsmath}	
\usepackage{pdflscape}
\usepackage{multirow}
\usepackage{makecell}
\usepackage{newtxtext,newtxmath}

\title[AGN fraction in galaxy pairs at cosmic noon]{Obscured AGN enhancement in galaxy pairs at cosmic noon: evidence from a probabilistic treatment of photometric redshifts}

\author[S. L. Dougherty et al.]{
Sean L. Dougherty,$^{1}$\thanks{E-mail: sldough21@gmail.com}
C. M. Harrison,$^{1}$\thanks{E-mail: christopher.harrison@newcastle.ac.uk}
Dale D. Kocevski$^{2}$ and
D. J. Rosario$^{1}$
\\
$^{1}$School of Mathematics, Statistics and Physics, Newcastle University, NE1 7RU, UK\\
$^{2}$Department of Physics and Astronomy, Colby College, Waterville, ME 04961, USA\\
}

\date{Accepted XXX. Received YYY; in original form ZZZ}

\pubyear{2023}

\begin{document}
\label{firstpage}
\pagerange{\pageref{firstpage}--\pageref{lastpage}}
\maketitle

\begin{abstract}
Observations of the nearby universe reveal an increasing fraction of active galactic nuclei (AGN) with decreasing projected separation for close galaxy pairs, relative to control galaxies. This implies galaxy interactions play a role in enhancing AGN activity. However, the picture at higher redshift is less established, partly due to limited spectroscopic redshifts. We combine spectroscopic surveys with photometric redshift probability distribution functions for galaxies in the CANDELS and COSMOS surveys, to produce the largest ever sample of galaxy pairs used in an AGN fraction calculation for cosmic noon ($0.5<z<3$). We present a new technique for assessing galaxy pair probability (based on line-of-sight velocities $\pm$1000\,km\,s$^{-1}$) from photometric redshift posterior convolutions and use these to produce weighted AGN fractions. Over projected separations 5--100\,kpc we find no evidence for enhancement, relative to isolated control galaxies, of X-ray ($L_{X}>10^{42}$\,erg\,s$^{-1}$) or infrared-selected AGN in major (mass ratios up to 4:1) or minor (4:1 to 10:1) galaxy pairs. However, defining the most obscured AGN as those detected in the infrared but not in X-rays, we observe a trend of increasing obscured AGN enhancement at decreasing separations. The peak enhancement, relative to isolated controls, is a factor of $2.08\pm0.61$ for separations $<$25\,kpc. Our simulations with mock data, indicates this could be a lower limit of the true enhancement. If confirmed with improved infrared imaging (e.g., with {\em JWST}) and redshifts (e.g., with forthcoming multi-object spectrograph surveys), this would suggest that galaxy interactions play a role in enhancing the most obscured black hole growth at cosmic noon. 

\end{abstract}




\section{Introduction} \label{sec:introduction}
There are now several observational results that indicate a co-evolution of supermassive black holes (SMBHs) and their host galaxies. This includes tight correlations between SMBH mass and spheroidal properties \citep[e.g.,][]{Magorrian98,Gebhardt00, Gultekin09,KormedyHo13} as well as the similar cosmic evolution of SMBH accretion and star formation rate (SFR) density, both of which peak at ``cosmic noon''  \citep[i.e., $z\sim2$; e.g.,][]{MadauDickinson14, Aird15}.

There is ongoing research on the role of galaxy mergers in driving the co-evolution of SMBHs and galaxies. Theoretically, strong gravitational interactions within a merger could reduce the angular momentum of gas and channel inflows into the nuclear region, both compressing gas into regions of intense star formation and accreting it onto the SMBH itself, consequently triggering an active galactic nuclei \citep[AGN; e.g.,][]{BarnesHernquist91, MihosHernquist96, BluementhalBarnes18}. Major galaxy mergers (i.e., those with mass ratios up to 4:1) undoubtedly impact the final morphology of the combined system \citep[e.g.,][]{Darg10, Ellison10, Casteels13}, but their implications for star formation \citep[e.g.,][]{Patton13, Barrera15, Moreno21, Ellison22} and AGN activity \citep[e.g.,][]{Treister12, Ellison13b, Villforth17, Goulding18} are a matter of significant debate in the literature. 

In the nearby universe ($z\sim0$), there is extensive observational evidence for mergers enhancing the number of AGN selected by optical emission lines \citep[][]{Ellison11, Ellison13b,Pierce23}, mid-infrared colours \citep[][]{Ellison13a, Satyapal14, Gao20} and those with high-excitation radio emission \citep[][]{RamosAlmeida11, Pierce22}. The largest excess of AGN associated with galaxy interactions or mergers is usually found for the most luminous sources; however, it is a matter of on-going debate if galaxy interactions significantly {\em increase} accretion rates \citep[][]{Ellison19,Pierce22,Bickley23,Steffen23}. These findings are well complemented in the cosmological hydrodynamic simulation \texttt{EAGLE}, where \citet{McAlpine20} report both the highest AGN excess in mergers at low redshifts ($0.05 < z < 0.1$) and the highest merging excess in AGN at high bolometric luminosities ($L_{\textrm{bol}} \sim 10^{45}\ \textrm{erg\,s}^{-1}$). 

This is all broadly in agreement with a scenario where, in the local Universe, the most luminous AGN phases can be boosted by major mergers whereas less luminous AGN are predominately associated with secular processes \citep[e.g.,][]{Sanders88a, Sanders88b, Bennert08, Urrutia08}. Nonetheless, the overall importance of galaxy mergers or interactions {\em for total black hole growth} remains unclear \citep[e.g., see discussion in][]{McAlpine20,ByrneMamahit22}. Secular processes, such as stochastic bar instabilities, stellar winds or interactions with dark matter halos, are alternate AGN fueling mechanisms for driving the majority of black hole growth \citep[e.g.,][]{KormendyKennicutt04, Hopkins06}.

Meanwhile, the AGN-merger connection is far from established at cosmic noon. Empirical works find conflicting results when starting from similar samples, such as luminous quasars \citep[][]{Gilkman15, Mechtley16, Marian19}, X-ray selected AGN \citep[][]{Cisternas11, Silverman11, Kocevski12, Lackner14, Rosario15, Hewlett17, Villforth17, Shah20} and infrared (IR) selected AGN \citep[][]{Schawinski12, Donley18, Goulding18, Silva21}. While a significant delay between the onset of inflows and AGN activity \citep[${\sim} 50{-}500$ Myrs; e.g.,][]{Davies07, Schawinksi09, Wild10} may make it difficult to reconcile contradicting findings, there are a couple of important cosmological factors to consider. First, the frequency of merging galaxies and luminous AGN is substantially higher at cosmic noon, relative to the local Universe \citep[${\sim} 10{\times}$; e.g.,][]{Conselice03, Kartaltepe07, Husko22}. Second, the gas fraction is similarly much greater \citep[${\sim} 5{\times}$; e.g.,][]{Daddi10, Tacconi10, Scoville14}, which could result in different inflow physics as well as more prominent phases of obscured black hole growth. These factors make a well considered control sample (of non-merging galaxies) crucial for any analyses of the AGN frequency.

Inconsistencies among works may, at least in part, be attributed to the various methods used to classify mergers. Most high-redshift studies rely on the identification of morphological disturbances indicative of mergers, such as tidal bridges or tails, in imaging data with model-fitting \citep[e.g.,][]{Lackner14}, machine learning \citep[e.g.,][]{Goulding18} or by visual inspection \citep[e.g.,][]{Cisternas11}. However, with a decrease in surface brightness $\propto (1+z)^{-4}$, this approach becomes exceptionally more challenging and temporally expensive when searching beyond nearby sources. This may lead to incorrect non-merging identifications and restrictive samples, particularly for the most luminous or obscured AGN.

An alternative approach, which is less susceptible to surface brightness dimming effects, is to look at close galaxy pairs and explore AGN fraction as a function of projected separation. This method has been used extensively at low redshift for emission-line- and IR-selected AGN \citep[e.g.,][]{Ellison11, Ellison13b, Satyapal14, Steffen23}. Recently, it has been applied at cosmic noon ($0.5 < z < 3.0$) by \citet{Shah20}, who used spectroscopic redshifts in the CANDELS and COSMOS fields to distinguish between random projections and truly associated galaxies. They find no excess in X-ray- or IR-selected AGN in close pairs compared to a matched control sample of isolated galaxies. This is in contrast to \citet{Silverman11}, who report mild excess of X-ray AGN (${\sim}2\times$) in $0.25 < z < 1.05$ spectroscopic galaxy pairs in COSMOS, albeit, at lower redshift range and when compared to a more general control sample (i.e., to the field). However, spectroscopic completeness at high redshift is poor due to source faintness and many strong emission lines being redshifted out of optical spectroscopy beyond $z \gtrsim 1.5$. Spectroscopic redshift incompleteness is particularly problematic for the most obscured sources (e.g., those not detected in X-ray surveys), which are much less frequently the target of spectroscopic observations. Without redshift information for {\em all galaxies}, true physical pairs remain inevitably entangled with chance projections. 

In this work, we address the issue of spectroscopic completeness at cosmic noon by folding in photometric redshift posterior probability distribution functions (PDFs). We build on previous studies of galaxy pairs and mergers that have utilised photometric redshifts \citep[e.g.,][]{Kartaltepe07, Bundy09, LS15, Mundy17, Mantha18, Duncan19}, but we develop a new approach, which is made novel by its fair treatment of the full PDF uncertainty.

The structure of this paper is as follows. In Section~\ref{sec:DataSampleSelection}, we describe the relevant archival data and define our parent sample. We outline our pair probability methodology in Section~\ref{sec:Method}. We present our results in Section~\ref{sec:Results} and discuss their implications for merger-driven SMBH growth at cosmic noon in Section~\ref{sec:Discussion}. Finally, we summarise our findings in Section~\ref{sec:Conclusion}. Throughout this work, we assume a $\Lambda$CDM cosmology with $H_0 = 70$ km\,s$^{-1}$ Mpc$^{-1}$, $\Omega_\Lambda = 0.7$ and $\Omega_M = 0.3$.

\section{Data and Sample Selection} \label{sec:DataSampleSelection}
Our study is designed to measure the prevalence of AGN as a function of galaxy separation, using a statistically robust sample of physically associated galaxy pairs and isolated control galaxies at cosmic noon ($0.5 < z < 3$). To achieve this, we have utilised the extragalactic deep fields from the Cosmic Assembly Near-infrared Deep Extragalactic Legacy Survey \citep[CANDELS;][]{Grogin11, Koekemoer11} and the Cosmic Evolution Survey \citep[COSMOS;][]{Scoville07}, following \citet{Shah20}. These fields have extensive multi-wavelength photometry, detailed archival analyses of photometric redshift probability distribution functions and stellar masses (Section~\ref{sec:PhotometryDerived}), and multiple spectroscopic redshift studies (Section~\ref{sec:Spectroscopy}). We use these as the basis for our parent galaxy sample selection (Section~\ref{sec:SampleSelection}) and for identifying both X-ray AGN and IR AGN (Section~\ref{sec:AGNselection}). 

\subsection{Photometry and derived quantities} \label{sec:PhotometryDerived}

We make use of the five CANDELS extragalactic deep fields, which were all observed with high-resolution near-IR and optical filters from the \textit{Hubble Space Telescope (HST)}/Wide-Field Camera 3 (WFC3) and \textit{HST}/Advanced Camera for Surveys (ACS). The CANDELS team has performed consistent deblending and photometry across fields, which combine for a total sky coverage of ${\sim}960$\,arcmin$^2$. To increase our source statistics, we also make use of the wider ${\sim} 2\ \textrm{deg}^2$ COSMOS field, the largest continuous area observed by {\em HST}/ACS to date. Below, we describe the archival ultra-violet (UV) to mid-infrared (MIR) photometry, the photometric redshift information, the derived stellar masses and the X-ray observations that we use in this study.   

\subsubsection{UV--MIR photometry}
The CANDELS photometry of the Great Observatories Origins Deep Survey North (GOODS-N) and South \citep[GOODS-S;][]{Giavalisco04}, Extended Groth Strip \citep[EGS;][]{Davis07}, UKIDSS Ultra Deep Survey \citep[UDS;][]{Lawrence07} and a central region from the COSMOS field are described in \citet{Barro19}, \citet{Guo13}, \citet{Stefanon17}, \citet{Galametz13} and \citet{Nayyeri17}, respectively. In short, the photometry is based on broadband data from the UV to MIR, including deep near-infrared (NIR) \textit{HST}/WFC3 and optical \textit{HST}/ACS CANDELS observations. Ancillary MIR observations were taken with \textit{Spitzer}/IRAC 3.4, 4.5, 5.8, and 8.0 $\mu$m imaging of GOODS-N \citep[][]{Dickinson03, Ashby15}, GOODS-S \citep[][]{Dickinson03, Ashby13}, EGS \citep[][]{Barmby08, Ashby13, Ashby15}, UDS \citep[][]{Ashby13, Ashby15} and COSMOS \citep[][]{Sanders07, Ashby13}. Source detection was performed with \texttt{SOURCE EXTRACTOR} \citep[][]{Bertins96} on the WFC3 F160W (\textit{H}-band) mosaic, which reaches $5\sigma$ limiting AB depths of 27.3-27.8 across the fields. Low-resolution photometry was measured with \texttt{TFIT} \citep[][]{Laidler07}. Briefly, this procedure involves convolving the brightness profile of high-resolution imaging (i.e., WFC3 F160W; \textit{H}-band) with a kernel to match the point-spread function of the low-resolution imaging (i.e., IRAC channels), from which sources can be deblended and fluxes extracted via best fit.

For the wider COSMOS field, we use the recent COSMOS2020 CLASSIC photometric catalog \citep[][]{Weaver22}. Once again, the photometry incorporates a wealth of multi-wavelength imaging from the UV-MIR, notably with recent optical Subaru/Suprime-Cam and NIR VISTA/ULTRAVISTA observations. In the MIR, IRAC observations were taken from the Cosmic Dawn Survey \citep[][]{Euclid22}, a collation of new and existing IRAC data in the Euclid deep fields, of which COSMOS is included \citep[][]{Sanders07, Ashby13, Ashby15, Ashby18, Steinhardt14}. Source detection was performed on a ``chi-squared'' $izYJHK_s$ image \citep[][]{Szalay99} with \texttt{SOURCE EXTRACTOR}. Low-resolution photometry was performed with \texttt{IRACLEAN} \citep[][]{Hsieh12}. Briefly, this method treats the high-resolution brightness profile as a prior, from which a fraction of its flux is iteratively subtracted until some minimal pixel brightness criteria is reached. Following \citet{Shah20}, we prioritise using the photometry and quantities derived by the CANDELS team in the region of COSMOS where the surveys overlap.

\subsubsection{Photometric redshift PDFs}
\label{Sec:PDFs}
In this study we utilise the full photometric redshift PDFs (in addition to spectroscopic redshift information) to calculate the probability of galaxies at close projected separations being in a physically associated pair (see Section~\ref{sec:Method}). That is, we do not simply assume the peak of the PDF as a fixed photometric redshift or its width as an uncertainty. Across all fields, these PDFs have been carefully constructed in previous works to both be accurate (i.e., the peak PDF estimate is similar to a confirmed spectroscopic redshift) and give realistic widths (i.e., $68\%$ of spectroscopic redshifts should fall within $1\sigma$ of the mean PDF estimate). 

For the wider COSMOS field, spectral energy distribution (SED) fitting was performed separately with \texttt{EAZY} \citep[][]{Brammer08} and \texttt{LePhare} \citep[][]{ArnoutsIlbert11} based on the COSMOS2020 photometry \citep[][]{Weaver22}. We elect to use the photometric redshift PDFs (and subsequent derived properties) from \texttt{LePhare} due to its greater precision across all magnitudes \citep[see Figure 15 in][]{Weaver22}. For the CANDELS fields, these PDFs have been independently generated from separate SED fitting codes with various template configurations by 6 groups in the CANDELS team. These codes are \texttt{EAZY} \citep[][]{Brammer08}, \texttt{zphot} \citep[][]{Giallongo98, Fontana00}, \texttt{HyperZ} \citep[][]{Bolzonella00}, \texttt{LePhare} \citep[][]{ArnoutsIlbert11} and \texttt{WikZ} \citep[][]{Wiklind08}. In an assessment of these methods, \citet{Dahlen13} find that the individual redshift posteriors underestimate realistic uncertainties and that they improve in accuracy when combined. Therefore, for each source, we use the PDF combination assembled by \citet{Kodra22} using a minimum Fre\'chet distance method, which was shown to be an improvement in both accuracy and PDF width over any individual group. We note that, despite the specific differences in PDF generation, we verified that our main results are consistent across CANDELS and COSMOS, if they are treated independently. 

We perform an assessment of the quality of the photometric redshifts in the context of the spectroscopic redshifts for the global galaxy population and the known AGN in Appendix~\ref{App:Spectroscopy}. In summary, we find that the normalized median absolute deviation between the spectroscopic redshifts and photometric redshifts is $\sigma_\textrm{NMAD} \sim 0.02$ and $\sigma_\textrm{NMAD} \sim 0.03$, for the full galaxy parent sample and only the AGN, respectively. Although these are modestly higher for the AGN, this is countered somewhat by the higher fraction of spectroscopic redshifts of the AGN compared to the general galaxy population (see Appendix~\ref{App:Spectroscopy}). We reiterate that the full PDF information is incorporated into our method (see Section~\ref{sec:Method}), and we further perform tests of including or removing spectroscopic redshifts in our analyses (see Section~\ref{sec:Full_Sample}).

\subsubsection{Stellar masses} \label{sec:masses}
We follow \citet{Shah20} in our choice of archival masses for the sample across fields used in this study. Specifically, in COSMOS, stellar masses were computed using \texttt{LePhare} (i.e., the same as that used for the photometric redshift PDFs) with the configuration described in \citet{Laigle16}. Medians of the photometric redshift PDFs generated by \texttt{LePhare} were taken as the fitting redshift. Stellar masses in the CANDELS fields were computed following the method outlined in \citet{Mobasher15} and \citet{Santini15}. In short, 10 groups within the CANDELS team estimated stellar mass PDFs from the same photometry and redshifts but with different SED stellar templates and/or codes. The median of the average stellar mass PDF is taken as the best mass estimate. 

We note that our results are insensitive to mass variations or uncertainties (e.g., caused by subtle differences in the chosen redshifts) because these are only used to create very broad categories of major and minor galaxy samples, for which we observe no differences in our results (see Section~\ref{sec:SampleSelection}).

\subsubsection{X-ray data} \label{sec:X-ray}
We make use of public \textit{Chandra} point-source catalogs to identify X-ray AGN in our sample. In GOODS-S, we use the point-source catalog of \citet{Xue11}. This benefits from the \textit{H}-band counterpart matching (directly to our preferred CANDELS photometric catalog) performed by \citet{Hsu14} and reaches observed X-ray luminosities down to $L_X = 10^{42}$\,erg\,s$^{-1}$ at $z \sim 3$, which is sufficient for our purposes. We use the 2\,Ms and 800\,ks point-source catalogs of \citet{Xue16} and \citet{Nandra15} for GOODS-N and EGS, respectively, and the optical counterparts therein. In COSMOS, we use the 4.6\,Ms catalog from \citet{Civano16} with optical counterparts identified in  \citet{Marchesi16}. Finally, we use the 600\,ks source catalog from \citet{Kocevski18} in UDS, matching the X-ray point-sources to the \textit{H}-band-selected objects in CANDELS using a maximum likelihood technique \citep[][]{SutherlandSaunders92, Civano12}.

\subsection{Spectroscopic redshifts} \label{sec:Spectroscopy}
We compile all publicly available secure spectroscopic redshifts in the CANDELS and COSMOS fields. Those spectroscopic redshifts were obtained from the measurements provided in several published works, which are detailed in full in Appendix~\ref{App:Spectroscopy}. 

There is no consistent quality flagging across these works; however, we select only spectroscopic redshifts measured from multiple emission lines. In the case where an object has multiple secure spectroscopic redshifts and the relative line-of-sight velocity calculated from the maximum difference in redshifts is less than ${\pm}1000$\,km\,s$^{-1}$ (i.e., the velocity offset used in our pair criteria; see Section~\ref{sec:Method}), we use the mean redshift. If this relative line-of-sight velocity is greater than ${\pm}1000$\,km\,s$^{-1}$, then we default to the spectroscopic redshift measured from the higher-resolution spectrograph. These final cases make up ${<}1\%$ of all spectroscopic redshifts used in this work. There are $32\%$ (${\sim}3500$) more spectroscopic redshifts in the parent sample (defined below) when including low-quality redshifts (i.e., those with single line identification or low signal-to-noise spectra). However, we prefer to focus on the high-quality redshifts only, using photometric redshifts otherwise (for which we take into account the full uncertainty). We note that our results are consistent if we only use photometric redshifts.

\subsection{Galaxy sample selection} \label{sec:SampleSelection} \label{ParentSample}
\begin{figure}
    \includegraphics[width=\columnwidth]{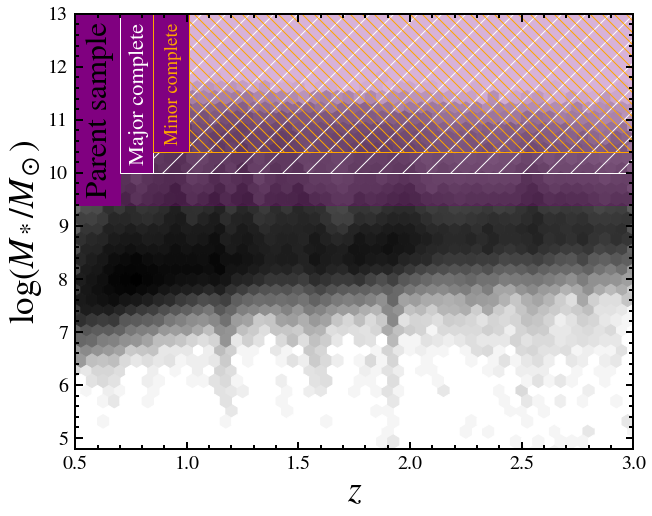}
    \caption{Stellar mass distribution of galaxies in the CANDELS and COSMOS fields. The shaded purple region, $M_* > 10^{9.4}\,M_\odot$, indicates all galaxies included in our parent sample. Fixing that galaxy 1 is always more massive than galaxy 2 (i.e., $M_{*,1} > M_{*,2}$), the white and orange striped regions define the stellar mass ranges of galaxy 1 to ensure a complete major pair sample ($M_* > 10^{10}\,M_\odot$ and $M_{*,1}{:}M_{*,2} < 4{:}1$) and minor pair sample ($M_* > 10^{10.4}\,M_\odot$ and $4{:}1 < M_{*,1}{:}M_{*,2} < 10{:}1$), respectively.}
    \label{fig:mass_sample}
\end{figure}

Our parent sample selection, primarily based on stellar mass and redshift, is presented in Figure~\ref{fig:mass_sample}. The selection is largely motivated by \citet{Shah20}, who also study galaxy pairs in the COSMOS and CANDELS fields. This allows for a more direct comparison of the X-ray and IR AGN between pairs identified with just spectroscopic redshifts (following \citealt{Shah20}) to a more complete sample, folding in the full photometric redshift PDFs for all galaxies, which is the procedure of this work (see Section~\ref{sec:Full_Sample}). 

Starting with the photometric catalogues described in Section~\ref{sec:PhotometryDerived}, we first clean the sample for poor photometric measurements. We exclude sources with a \texttt{SOURCE EXTRACTOR} stellarity parameter greater than 0.9 to account for high-probability stellar contaminants \citep[see][]{Guo13}, and we remove sources where the photometry is compromised by image artifacts, such as edge effects, nearby bright objects or diffraction spikes. 

We then select the parent sample as all galaxies with $M_* > 10^{9.4} M_\odot$ and $0.5 < z < 3$ (purple shaded region in Figure~\ref{fig:mass_sample}). Since we are not limited to sources with spectroscopic redshifts, our sample is large enough to separately investigate AGN activity in both major and minor mergers, which we define as pairs with mass ratio $M_{*,1}{:}M_{*,2} < 4{:}1$ and $4{:}1 < M_{*,1}{:}M_{*,2} < 10{:}1$, respectively. Designating $M_{*,1} > M_{*,2}$, our parent sample is complete for major mergers when $M_{*,1} > 10^{10} M_\odot$ and for minor mergers when $M_{*,1} > 10^{10.4} M_\odot$ (see white and orange striped regions in Figure~\ref{fig:mass_sample}). Consequently, we define two pair samples, major complete ($M_{*,1} > 10^{10} M_\odot$ and $M_{*,1}{:}M_{*,2} < 4{:}1$; i.e., most similar to the parent sample of \citealt{Shah20}) and minor complete ($M_{*,1} > 10^{10.4} M_\odot$ and $4{:}1 < M_{*,1}{:}M_{*,2} < 10{:}1$).

\subsection{AGN selection} \label{sec:AGNselection}
\begin{figure}
    \includegraphics[width=\columnwidth]{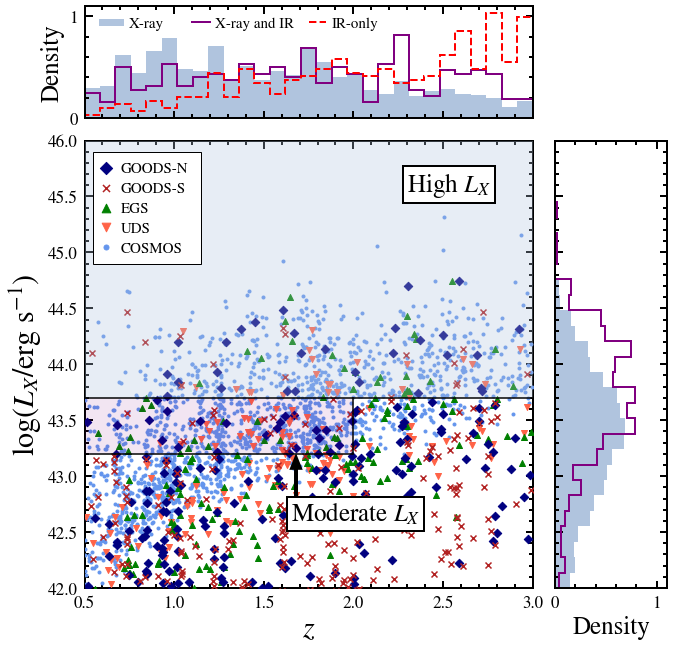}
    \caption{X-ray luminosity (full band, 0.5-10\,keV) versus redshift for all X-ray sources ($L_X > 10^{42}$\,erg\,s$^{-1}$) in our parent sample. The navy diamonds, red crosses, green upward triangles, orange downward triangles and blue dots correspond to X-ray sources in GOODS-N, GOODS-S, EGS, UDS and COSMOS, respectively. We define two large $L_X$ regimes that are complete across all fields. The purple shaded region corresponds to ``moderate $L_X$ AGN'' ($L_X = 10^{43.2{-}43.7}$\,erg\,s$^{-1}$), with $0.5 < z < 2$. The blue shaded region defines the ``high $L_X$ AGN'' sample ($L_X > 10^{43.7}$\,erg\,s$^{-1}$), over the full redshift range. Redshift and luminosity distributions are shown in the top and right panels, respectively, for AGN sub-samples: AGN selected in the X-ray (solid blue histogram); those also selected in X-ray and IR (purple line) and those selected only in the IR (i.e., $L_X < 10^{42}$\,erg\,s$^{-1}$ or undetected; red dashed line).}
    \label{fig:LX_sample}
\end{figure}

X-ray fluxes in the 0.5-10\,keV band were extrapolated from the observed fluxes in the full (0.5-8\,keV), hard (2-8\,keV), and soft (0.5-2\,keV) bands (in that order of priority) assuming an X-ray spectral index of slope $\Gamma = 1.4$ \citep[e.g.,][]{Marchesi16}. The total X-ray luminosity, $L_X$, is then
\begin{equation} \label{eq:LX}
    L_X = F_X \times 4\pi d_L^2 \times (1+z)^{\Gamma-2},
\end{equation}
where $F_X$ is the extrapolated X-ray flux and $d_L$ the luminosity distance (calculated using the peak of the PDF photometric redshift or spectroscopic redshift, if available). We then impose a conservative AGN threshold cut of $L_X > 10^{42}$\,erg\,s$^{-1}$ \citep[e.g.,][]{Moran99} to avoid any contamination from high-mass X-ray binary emission in star-forming galaxies \citep[][]{Alexander05}. These AGN are represented in an X-ray luminosity-redshift plane in Figure~\ref{fig:LX_sample}. We note that there is some uncertainty in the X-ray luminosities, particularly due to the photometric redshifts; however, these do not affect our results. The luminosities are only used to define broad X-ray luminosity bins, and, when modifying the X-ray luminosity threshold, our results remain quantitatively and qualitatively consistent (see Section~\ref{sec:Complete_Xray}). 

We define two X-ray AGN samples, with different X-ray luminosity ranges, to reflect the varying depths of the {\em Chandra} observations across the CANDELS and COSMOS fields. We define ``moderate $L_X$ AGN'' as those with $L_X = 10^{43.2{-}43.7}$\,erg\,s$^{-1}$, complete across all fields to $0.5 < z < 2$ (see purple shaded region in Figure~\ref{fig:LX_sample}), and ``high $L_X$ AGN'' with $L_X > 10^{43.7}$\,erg\,s$^{-1}$, complete over the full redshift range  (see blue shaded region in Figure~\ref{fig:LX_sample}).

Following \citet{Shah20}, we classify IR AGN from IRAC data based on the criteria outlined in \citet{Donley12}. This method combines a colour-colour cut with a power law selection, excluding star-forming galaxies at this epoch and assembling a highly complete and reliable sample of IR AGN. Other IRAC selection methods just use a simple colour-colour cut to identify IR AGN \citep[][]{Lacy04, Lacy07, Stern05}, but they become contaminated by star-forming galaxies at $z \gtrsim 0.5$ \citep[e.g.,][]{Assef10, Park10} and are therefore less conservative.

We note that $14.6\%$ of X-ray-selected AGN are also IR-selected, but $47.7\%$ of IR-selected AGN are not X-ray selected as AGN. We investigate this latter population further in Section~\ref{sec:Obscured_AGN}. Additionally, given the propensity of X-ray-selected AGN for spectroscopic follow-up, $30\%$ of X-ray-selected AGN have a secure spectroscopic redshift compared to ($8.3\%$) $21\%$ of AGN selected (only) in the IR. The redshift distribution (peak photometric estimate or secure spectroscopic redshift, if available) of X-ray AGN, those also selected in the IR and IR-only AGN are characterised in Figure~\ref{fig:LX_sample}. Photometric PDF validation is performed for these sub-samples using spectroscopic redshifts in Appendix~\ref{app:phot_val}.

\section{Probabilistic pair methodology} \label{sec:Method}
In this section, we outline the method used to assess the probability that a projected pair of galaxies is in a physically associated ``true'' pair and how we construct a control sample of isolated galaxies. We use this information to calculate the AGN fraction and AGN enhancement (with respect to the control galaxies) as a function of projected separation. 
We define a true galaxy pair as two galaxies within projected separation $r_p < 100$\,kpc and a relative line-of-sight velocity $|\Delta V| < 1000$\,km\,s$^{-1}$. Our choice of $|\Delta V| < 1000$\,km\,s$^{-1}$, to define physically associated pairs, follows that chosen for the primary analyses of \citet{Shah20}. We also note that enhanced AGN fractions are typically found to start in pairs of $r_p \lesssim 50$\,kpc \citep[e.g.,][]{Ellison13b, Satyapal14}. 

Broadly, our approach is as follows. We combine photometric redshift PDFs to compute, for each candidate galaxy pair, the probability that $|\Delta V|$ is within 1000\,km\,s$^{-1}$ ($\mathcal{P}_{\Delta V}$; see Section~\ref{sec:P_DV}) and that their projected separation falls within each separation bin out to $r_p = 100$\,kpc ($\mathcal{P}_r^\textrm{bin}$; Section~\ref{sec:P_r}). Combining $\mathcal{P}_{\Delta V}$ and $\mathcal{P}_r^\textrm{bin}$, we compute the true pair probability for each pair in each projected separation bin ($\mathcal{P}_\textrm{pair}^\textrm{bin}$; Section~\ref{sec:P_pair}). We then carefully assemble a control sample of physically unassociated projected pairs, matched in redshift, stellar mass, environmental density and redshift quality (Section~\ref{sec:Control}). Finally, we calculate and compare weighted AGN fractions in the true pair and control samples (Section~\ref{sec:AGNfracEnh}).

\subsection{Relative line-of-sight probabilities} \label{sec:P_DV}
We begin by assembling a pool of candidate projected galaxy pairs, defined as those separated by less than $16.4$\,arcsec. This distance was chosen because it roughly corresponds to a projected separation of 100\,kpc at the lowest redshift of the parent sample ($z=0.5$). This step ensures that we obtain all pairs with $r_p < 100$\,kpc across the full redshift range. Due to our large parameter space and the density of sources in these deep fields, a galaxy has anywhere from 0 to 36 candidate projected pairs in our parent sample within 16\,arcsec, all of which we assign a ``true pair'' probability (described below).

Once we have candidate pairs, we assess the probability that their relative line-of-sight velocity is within 1000\,km\,s$^{-1}$, $\mathcal{P}_{\Delta V}$. This approach relies on careful incorporation of the photometric redshift PDFs, resulting in the galaxy pairs with wider PDFs receiving lower pair probabilities. 

We perform our analysis with and without the available spectroscopic redshifts (see Section~\ref{sec:Full_Sample}), although we include the spectroscopic information as our primary approach. There are three possible pair types: (1) where both galaxies have a spectroscopic redshift ($z_{\textrm{spec}}$); (2) both have photometric redshift PDFs; and (3) one of each redshift type. Calculating $\Delta V$ and $r_p$ is trivial for type (1), and we assign $\mathcal{P}_{\Delta V} = 0$ or $\mathcal{P}_{\Delta V} = 1$ for these cases, which make up ${\sim}1.5\%$ of all initial pair candidates. For evaluating cases (2) and (3), we introduce a probability convolution technique to combine the redshift probability distributions and create a relative line-of-sight velocity probability distribution function, $P(\Delta V)$. We interpret a spectroscopic redshift as a dirac delta distribution, $\delta(z_{\textrm{spec}})$. Photometric redshift PDFs and $\delta(z_{\textrm{spec}})$ are henceforth both included in ``$P(z)$,'' as they are treated equally in the methodology. 

As we define a true pair based on a $\Delta V$ threshold, we begin with a simple change of variables from redshift to line-of-sight velocity space, $P(z) \rightarrow P(V)$, for galaxies 1 and 2 in a pair. We convolve $P_1(V)$ with $P_2(V)$ to obtain the probability distribution function of their relative line-of-sight velocity, $P(\Delta V)$, following:
\begin{equation} \label{eq:convolution}
    P(\Delta V) = \int P_1(V_{1}=V) \ P_2(V_{2}=V + \Delta V) \ dV.
\end{equation}
We then integrate this distribution within our chosen $\Delta V$ threshold ([-1000, 1000]\,km\,s$^{-1}$) to obtain $\mathcal{P}_{\Delta V}$:
\begin{equation} \label{eq:PdV}
    \mathcal{P}_{\Delta V} = \int_{-1000\ \textrm{km\,s}^{-1}}^{1000\ \textrm{km\,s}^{-1}} P(\Delta V) \ d\Delta V.
\end{equation}

\begin{figure*}
    \includegraphics[width=\textwidth]{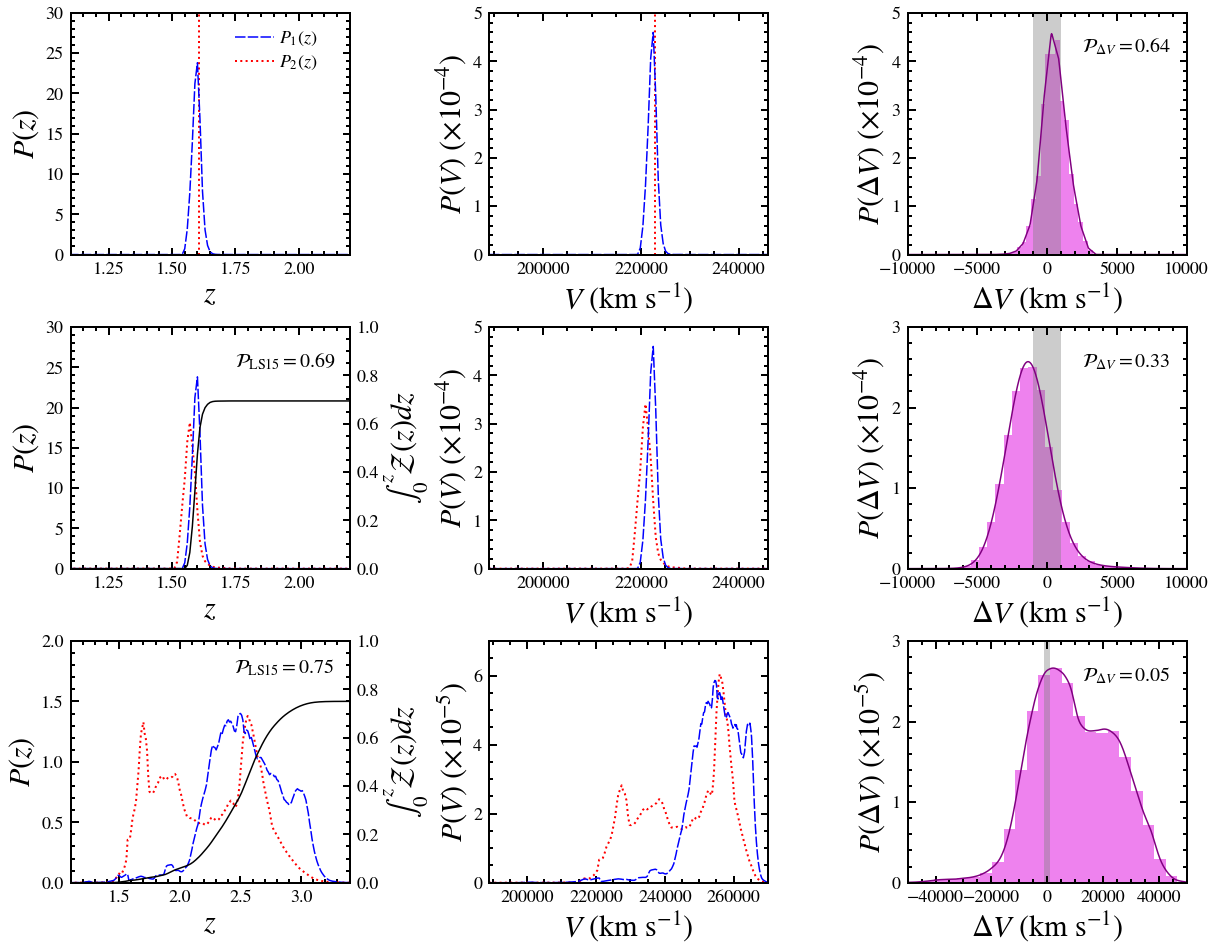}
    \caption{Determination of the relative line-of-sight probability, $\mathcal{P}_{\Delta V}$, for three example projected pairs (one in each row). From top to bottom: {\em Row 1:} the galaxy pair includes one spectroscopic redshift and one good quality photometric redshift PDF; {\em Row 2:} two good quality photometric PDFs; {\em Row 3:} two poor quality PDFs. \textit{Left column}: Overlaid redshift PDFs, $P_1(z)$ (dashed curve) and $P_2(z)$ (dotted curve), where a vertical line corresponds to a spectroscopic redshift. The black curve in the middle and bottom panels is the cumulative integral of the combined PDFs, $\mathcal{P}_\textrm{LS15}$, which has been used in previous work to estimate true pair probability (see Section~\ref{sec:Disc-PDF}). \textit{Middle column}: line-of-sight velocity distributions. \textit{Right column}: relative line-of-sight velocity distributions computed from Equation~\ref{eq:convolution} (solid curve). The purple histogram shows the $\Delta V$ distribution from a Monte Carlo simulation, verifying the accuracy of the convolution method. The vertical shaded region indicates the integration bounds of $\mathcal{P}_{\Delta V}$ for our true pair definition ($|\Delta V| < 1000$\,km\,s$^{-1}$).}
    \label{fig:conv_method}
\end{figure*}

In Figure~\ref{fig:conv_method} we provide a visual representation of the probability convolution for the following examples: a high probability case involving a photometric redshift PDF and a $z_\textrm{spec}$ (top row); a moderate probability case of two PDFs (middle row); and a low probability case of two PDFs (bottom row). While $\mathcal{P}_{\Delta V}$ increases when the means of the two PDFs are similar, it is appropriately decreased by the individual PDF widths. As a result, we do not need to exclude sources with large redshift uncertainties or correct for chance pairs as is sometimes applied \citep[e.g.,][]{Bundy09, Mantha18}. Furthermore, our method penalises against very broad PDFs regardless of how similar their means are, which is not the case for combined redshift probability methods based purely on PDF overlap \citep[e.g.,][see Section~\ref{sec:Disc-PDF}]{LS15, Duncan19}.

We validate our convolutional pair probabilities with a ``brute-force'' Monte Carlo method, where we draw redshifts from their underlying PDFs, 10,000 times\footnote{A $z_\textrm{spec}$ is drawn each iteration when available for the supplemented sample.}, with which we calculate $\Delta V$ each iteration. These distributions of $\Delta V$ (shown as filled histograms) are plotted together with the probability convolutions (shown as curves) in the rightmost panels of Figure~\ref{fig:conv_method}, demonstrating their excellent agreement. 

\subsection{Projected separation probabilities} \label{sec:P_r}

\begin{figure*}
    \includegraphics[width=\textwidth]{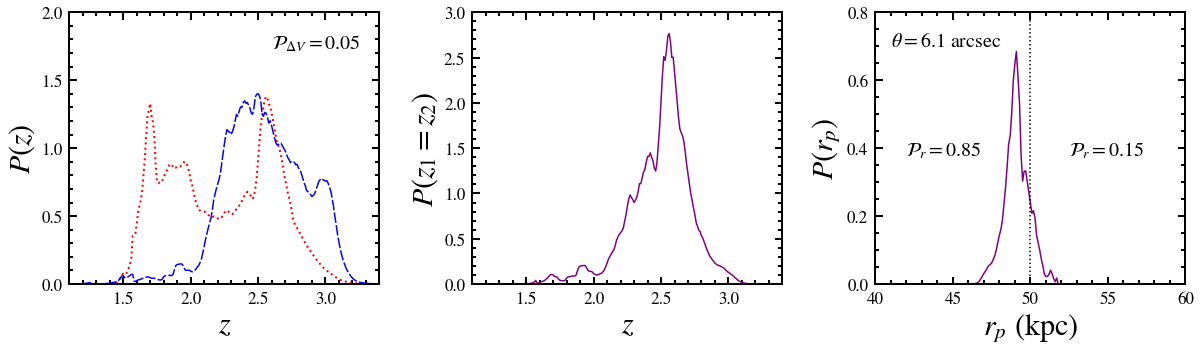}
    \caption{Example to show the method of calculating the projected separation probability, $\mathcal{P}_r$, using the pair shown in the bottom row of Figure~\ref{fig:conv_method}, which has an angular projected separation of 6.1\,arcsec. \textit{Left panel}: overlaid photometric redshift PDFs, $P_1(z)$ and $P_2(z)$. \textit{Middle panel}: combined redshift distribution (Equation~\ref{eq:Pz1z2}). \textit{Right panel}: projected separation distribution, $P(r_p)$, obtained from a change of variables of the combined redshift distribution. In this case, there is a probability of $\mathcal{P}_r=0.85$ that this pair falls in the projected separation bin of $25{-}50$\,kpc and a probability of $\mathcal{P}_r=0.15$ that it falls in the $50{-}75$\,kpc bin.}
    \label{fig:PdA}
\end{figure*}

The projected separation in physical units of a galaxy pair, $r_p$, is also dependent on redshift (i.e., the conversion from arcseconds to kiloparsecs). Therefore, we require the uncertainties in $P(z)$ to be reflected in our determination of $r_p$ as well, specifically when we wish to assign a galaxy pair into a particular bin of $r_p$ to investigate trends with projected separation as is commonly done \citep[e.g.,][]{Ellison13b, Shah20}. Following the logic of the previous section, we assess the probability, $\mathcal{P}_{r}$, a pair is within some projected separation, $r_p^\textrm{min}$ and $r_p^\textrm{max}$, from a projected separation probability distribution function, $P(r_p)$. First, we combine $P_1(z)$ and $P_2(z)$ to obtain a distribution of how likely they are to both be at a certain redshift:
\begin{equation} \label{eq:Pz1z2}
    P(z_1=z_2) = P_1(z) \times P_2(z).
\end{equation}
In other words, if the two galaxies were truly associated, this distribution tells us what that shared redshift likely would be. We normalize and perform a change of variables on $P(z_1=z_2)$ to arrive at $P(r_p)$, which we evaluate between $r_p^\textrm{min}$ and $r_p^\textrm{max}$ to obtain the projected separation probability:
\begin{equation} \label{eq:rProb}
    \mathcal{P}_r = \int_{r_p^{\textrm{min}}}^{r_p^{\textrm{max}}} P(r) \ dr.
\end{equation}
This process is illustrated in Figure~\ref{fig:PdA} for the low probability case (bottom row) of Figure~\ref{fig:conv_method}.

To compare samples versus $r_p$ in a statistically significant way, we sort our pairs into $r_p$ bins of width 25 kpc. To limit the impact of source confusion, we exclude pairs with separations of less than 5\,kpc, following previous work on high-redshift pair finding studies \citep[e.g.,][]{Mantha18, Duncan19}. The bin boundaries become the integration limits $r_p^\textrm{min}$ and $r_p^\textrm{max}$ in Equation~\ref{eq:rProb}, from which we determine $\mathcal{P}_r$ for all 4 bin regions (beginning with $\mathcal{P}_r^\textrm{5-25 kpc}$). This approach ensures that arbitrary bin selection does not misappropriate projected separations estimated from uncertain redshifts.

\subsection{True pair probability} \label{sec:P_pair}
To obtain a final true pair probability for a candidate galaxy pair in each projected separation bin, $\mathcal{P}_{\textrm{pair}}^{\textrm{bin}}$, we combine the relative line-of-sight probability ($\mathcal{P}_{\Delta V}$) and projected separation probability ($\mathcal{P}_r$), following:
\begin{equation}\label{eq:trueP}
    \mathcal{P}_{\textrm{pair}}^{\textrm{bin}} = \mathcal{P}_{\Delta V} \times  \mathcal{P}_r^{\textrm{bin}}.
\end{equation}
Continuing the pair example shown in both Figure~\ref{fig:conv_method} (bottom row) and Figure~\ref{fig:PdA}, with an angular separation of 6.1\,arcsec, we calculate $\mathcal{P}_\textrm{pair}^\textrm{25-50\,kpc} = 0.05 \times 0.85 = 0.0425$ and $\mathcal{P}_\textrm{pair}^\textrm{50-75\,kpc} = 0.05 \times 0.15 = 0.0075$. We use these pair probabilities as weights to calculate a weighted AGN fraction (see Section~\ref{sec:AGNfracEnh}), where more confident pairs (i.e., those with higher $\mathcal{P}_\textrm{pair}$) contribute more, following a similar weighting method previously used by \citet{Duncan19}. 

\subsection{Control selection} \label{sec:Control}
To achieve our goal of establishing the effect of close galaxy pairs on the prevalence of AGN, we need to compare the AGN frequency in our true pair sample to that of a physically unassociated isolated sample controlled for mass, redshift and environmental density \citep[e.g.,][]{Ellison13b, Shah20}. These three properties all may play a role in the rate of AGN, though they may not be directly related to close galaxy pair separation \citep[e.g.,][]{Perez09}. Following \citet{Ellison13b}, we define environmental density as the total number of sources within $|\Delta V| < 1000$\,km\,s$^{-1}$  and $r_p < 1$\,Mpc over the corresponding projected unit area (calculated using our probabilistic method; see Appendix~\ref{app:control_selection}). However, we note that this has $10\times$ lower weighting in the matching criteria compared to mass, redshift and PDF quality. As we interpret true pair probabilities as weights in an ultimate weighted AGN fraction calculation (Section~\ref{sec:AGNfracEnh}), we must ensure that any biases associated with PDF quality (i.e., $P(z)$ width) are controlled for as well.

Generally, it is trivial to find several matches for each paired galaxy in the often large pool of isolated galaxies, but ${\sim}47\%$ of all galaxies in our parameter space have at least one projected companion with a $\mathcal{P}_{\Delta V} > 0.01$. That is, around half of the galaxies have at least one other galaxy, within 100\,kpc, with a projected pair probability greater than 0.01. Since our definition of a true pair isn't binary but rather continuous (i.e., $\mathcal{P_\textrm{pair}^\textrm{bin}}$ is between 0 and 1), we must treat isolated galaxies as such. 

\begin{figure*}
    \includegraphics[width=\textwidth]{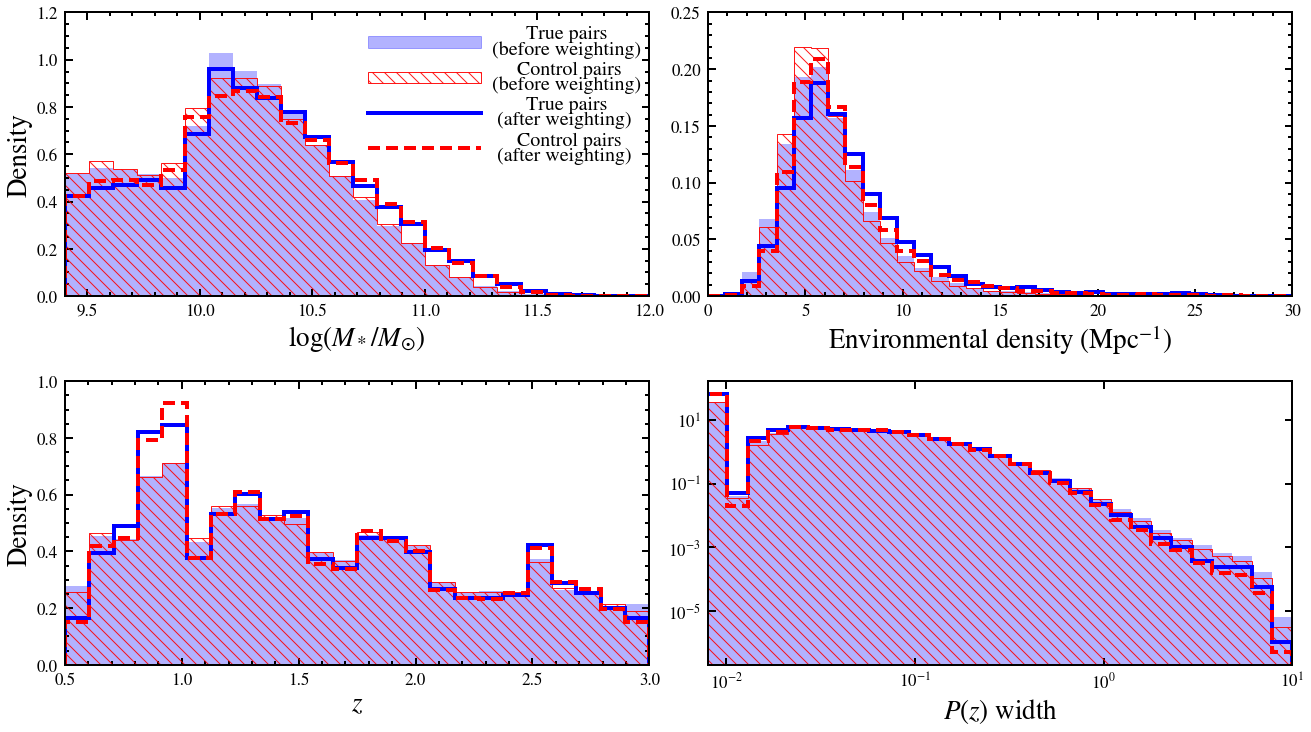}
    \caption{Distributions of stellar mass (\textit{top left}), redshift (\textit{bottom left}), environmental density (\textit{top right}) and $P(z)$ widths (\textit{bottom right}) for all true pairs and isolated controls. The shaded blue and striped red histograms correspond to the true and control distributions without weighting by pair probability, $\mathcal{P}_\textrm{pair}$. The solid blue and dashed red step histograms show the true and control pair distributions after applying the weighting used in the analyses, which demonstrates that the controls are well matched in these properties.}
    \label{fig:control_match}
\end{figure*}

Our full probabilistic approach to selecting control galaxies is supplied in detail in Appendix~\ref{app:control_selection}. Briefly, for each true pair (TP), we select three control pairs (CP) of physically unassociated galaxies (six unique galaxies; i.e., $\mathcal{P}_r^{0{-}100} = 0$) matched in mass, redshift, environment and $P(z)$ width. The three CPs are matched to the relative line-of-sight probability of the TP to ensure statistically comparable weighting (Section~\ref{sec:AGNfracEnh}). Normalized distributions of the matched parameters with and without weights are shown in Figure~\ref{fig:control_match}. We report that more than $90\%$ of control pairs are matched to within stellar mass of 0.2 dex, redshift of 0.2, environmental density of 4 and $P(z)$ width of 0.2 dex of their true pair counterparts (this includes to both galaxy 1 and 2) when $\mathcal{P}_{\Delta V}^\textrm{TP} > 0.01$. This confirms that our control matching process has been successful and that the effects of galaxy proximity on AGN frequency will be isolated in our analysis. 

\subsection{Weighted AGN fraction and AGN enhancement} \label{sec:AGNfracEnh}
With a sample of true pairs and a matched sample of unassociated controls (Section~\ref{sec:Control}), we can analyze the effects of galaxy separation on AGN fraction. In Sections~\ref{sec:P_r}, we split our true pair sample into four projected separation bins of width 25 kpc by integrating Equation~\ref{eq:rProb} between the bin edges. In Section~\ref{sec:P_pair}, we combine these bin probabilities, $\mathcal{P}_r^\textrm{bin}$, with the relative line-of-sight probability of each true pair, $\mathcal{P}_{\Delta V}$, to obtain the true pair probability for each bin, $\mathcal{P}_\textrm{pair}^\textrm{bin}$ (Equation~\ref{eq:trueP}). Here, we quantify the frequency of AGN of each bin with a weighted AGN fraction, defined as
\begin{equation} \label{eq:weightedAGNfrac}
    \textrm{Weighted AGN Fraction} = \frac{\sum_i (\mathcal{P}_{\textrm{pair},i}^\textrm{bin} \times N_{\textrm{AGN},i})}{\sum_i \mathcal{P}_{\textrm{pair},i}^\textrm{bin}},
\end{equation}
where $N_{\textrm{AGN},i}$ is the number of AGN (0, 1 or 2) in pair $i$. Rather than counting AGN in all paired galaxies, we count AGN in galaxy pairs weighted by how likely they are to be truly associated. In other words, the weighted AGN fraction in each pair is the weighted mean of AGN counts per pair and ranges from 0 to 2. In order to isolate and quantify the effects of galaxy separation on AGN activity, we need a weighted AGN fraction of the control pairs, which we calculate from Equation~\ref{eq:weightedAGNfrac} by replacing $\mathcal{P}_\textrm{pair}^\textrm{bin}$ with the control relative line-of-sight probability multiplied by the corresponding true pair projected separation probability. Multiplying $\mathcal{P}_{\Delta V}^\textrm{CP}$ by $\mathcal{P}_r^\textrm{bin}$ ensures, for instance, that a true pair of $\mathcal{P}_{\Delta V}^\textrm{TP}$ split between the 25-50 and 50-75 kpc bins will have its matched control pairs of $\mathcal{P}_{\Delta V}^\textrm{CP} \approx \mathcal{P}_{\Delta V}^\textrm{TP}$ shared between bins in the same way (Figure~\ref{fig:PdA}).

Finally, we define AGN enhancement (also known as AGN excess) as the ratio of the weighted AGN fraction for the true and control pairs of each bin:
\begin{equation} \label{eq:AGN_enh}
    \textrm{AGN Enhancement} = \frac{\textrm{Weighted AGN Fraction}_{\text{TP}}}{\textrm{Weighted AGN Fraction}_{\text{CP}}}.
\end{equation}
We use a common bootstrap technique \citep{Efron79, Efron81} to estimate the standard error for the weighted AGN fractions, which we propagate to estimate errors on AGN enhancement.

In Appendix~\ref{app:simulation} we apply this method, using only photometric redshifts, to a mock sample of galaxies with a true AGN enhancement. Although it is very difficult to capture the full complexities of the datasets, and caution should be taken in over-interpreting the quantitative simulated results, the simulation aims to roughly mimic the quality of the PDFs used in this work. We find that even with only photometric redshifts, we are able to recover an increase in the weighted AGN fraction of close galaxy pairs. The measured AGN fractions may be moderately underestimated on our main result presented for the obscured AGN in Section~\ref{sec:Obscured_AGN}; however, the large error bars on this final result capture some of this uncertainty on the exact level of enhancement. Only with a large number of spectroscopic redshifts would we be able to provide a more precise final measurement.

\section{Results} \label{sec:Results}
In this section, we report weighted AGN fractions (i.e., the weighted average AGN count per single galaxy pair) and AGN enhancements (relative to matched control galaxies), as defined in Section~\ref{sec:AGNfracEnh}. We show this  separately for X-ray and IR AGN samples, and we investigate any trends in these quantities as a function of projected separation (Section~\ref{sec:Full_Sample}). We go on to define sub-samples to explore any additional trends in these results with X-ray luminosity or redshift (Section~\ref{sec:Complete_Xray}). Finally we investigate a sample of AGN which are only identified in the IR but not selected in the X-rays (Section~\ref{sec:Obscured_AGN}). Our main results incorporate spectroscopic redshifts; however, we present some results using solely photometric redshift PDFs for validation (shown as grey data points in Figure~\ref{fig:XAGN_IRAGN}). All data plotted is tabulated in Appendix~\ref{app:tables}.

\subsection{Full sample} \label{sec:Full_Sample}

\begin{figure*}
    \includegraphics[width=\textwidth]{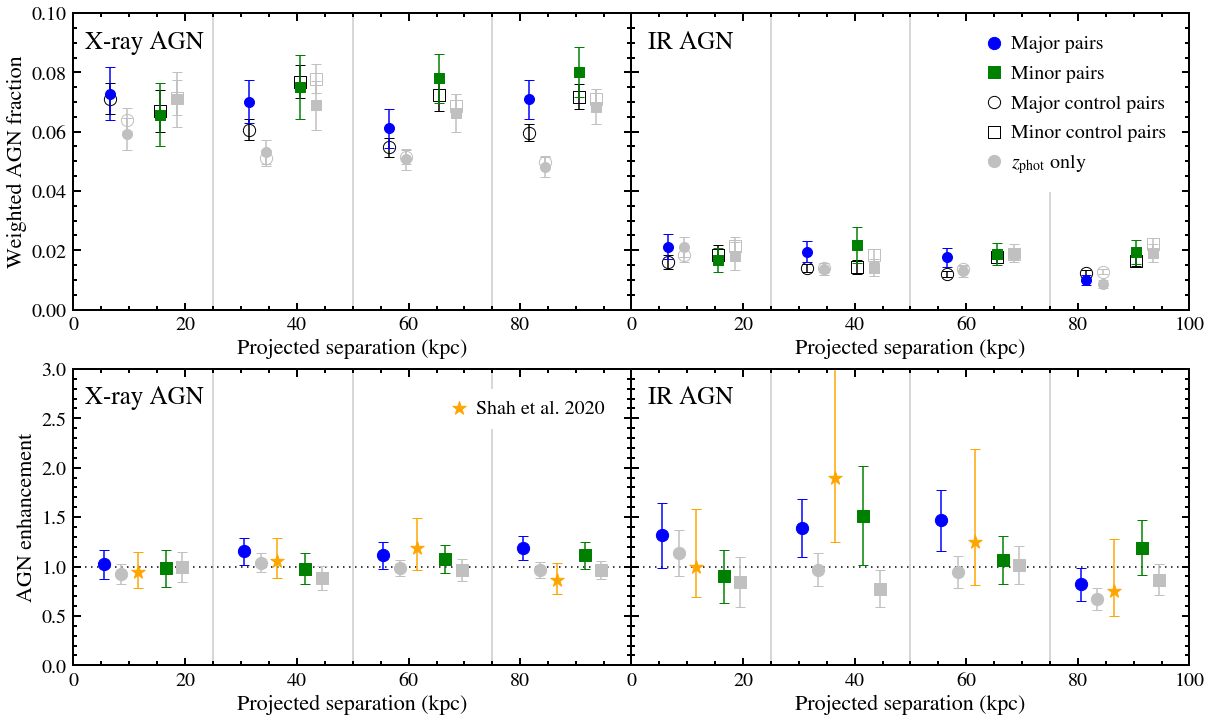}
    \caption{Results of weighted AGN fractions ({\em top row}) and AGN enhancement (relative to the controls; {\em bottom row}) as a function of projected separation. AGN fractions and enhancements for all $0.5 < z < 3$ X-ray AGN and IR-selected AGN are shown on the left and right, respectively. In all panels, the filled blue circles and filled green squares indicate major and minor galaxy true pairs, respectively, as defined in Figure~\ref{fig:mass_sample}.  Open circles and squares indicate the weighted AGN fractions for the corresponding control samples. Grey points show the same weighted AGN fractions found without the inclusion of spectroscopic redshifts, denoted as $z_\textrm{phot}$ only. Also plotted are AGN enhancements of major galaxy pairs from \citet{Shah20}, who only used spectroscopic redshifts in the same fields. Vertical grey lines indicate the projected separation bin edges. Data points are horizontally offset for visual aid. The dotted horizontal lines correspond to an AGN enhancement of 1; all data points scatter about this line, indicating no enhancement of AGN in true galaxy pairs relative to the control.}
    \label{fig:XAGN_IRAGN}
\end{figure*}

The weighted AGN fractions in bins of projected separation for all X-ray AGN ($L_X > 10^{42}$\,erg\,s$^{-1}$, $0.5 < z < 3$) are presented for our major-complete and minor-complete samples in the top-left panel of Figure~\ref{fig:XAGN_IRAGN}. For both the major and minor samples (filled coloured points) we see no trend of weighted AGN fraction as a function of projected separation, with all values scattering around ${\sim}0.07$. The control galaxies (hollow points) are always in good agreement with the true pairs. Alongside the primary data points, which include both spectroscopic and photometric information, we show the results from the purely photometric redshift approach, denoted ``$z_\textrm{phot}$ only'' (not using $\delta(z_\textrm{spec})$). We remind the reader that spectroscopic surveys also specifically target X-ray sources, which can boost the spectrosopic redshift fraction (see Section~\ref{sec:AGNselection}) and thus the chances of X-ray sources being detected in a pair. Although the weighted AGN fraction values are slightly lower ($\sim$0.055) for the major galaxy pair sample when spectroscopic redshifts are not used, this decrease is matched by the corresponding control galaxy pairs.

X-ray AGN enhancement, that is the ratio of the weighted AGN fractions of the true and control pairs, is plotted for these samples in the bottom-left panel of Figure~\ref{fig:XAGN_IRAGN}. The dotted line at an AGN enhancement value of 1 illustrates the case where the weighted AGN fractions are the same for both the true pairs and controls (i.e., no enhancement). Both major and minor pairs exhibit no evidence for AGN enhancement at small projected separations ($r_p < 25$\,kpc), with values of $1.02 \pm 0.15$ and $0.98 \pm 0.19$, respectively. This finding holds regardless of the inclusion of spectroscopic redshifts, with the $z_\textrm{phot}$ only method resulting in enhancements of $0.93 \pm 0.10$ and $0.99 \pm 0.15$ for major and minor galaxy pairs, respectively. Additionally, this result is consistent with \citet{Shah20}, who find no X-ray AGN enhancement for major pairs while only using spectroscopic redshifts for the same fields (shown with yellow stars in Figure~\ref{fig:XAGN_IRAGN}).

We plot the corresponding results for all IR AGN \citep[following][$0.5 < z < 3$]{Donley12} in the right panels of Figure~\ref{fig:XAGN_IRAGN}. Given the conservative selection criteria, there are ${\sim}4\times$ fewer IR AGN than X-ray AGN. As a result, IR AGN enhancements have larger errors. We find no statistically significant AGN enhancement with decreasing projected separation. For the smallest separations ($r_p < 25$\,kpc), the enhancement values are $1.32 \pm 0.33$ in major and $0.90 \pm 0.27$ in minor pairs. Once again, these results are consistent to those of the $z_\textrm{phot}$ approach, which results in enhancements of $1.14 \pm 0.23$ and $0.84 \pm 0.25$ for major and minor galaxy pairs, respectively, at the smallest separations. These findings are consistent with the IR AGN enhancements reported by \citet{Shah20} in $0.5 < z < 3$ major galaxy pairs using only spectroscopic redshifts, albeit their error bars are significantly larger due to lower number statistics.

The results presented in the following two subsections focus on major galaxy pairs with the incorporation of spectroscopic redshifts where available.

\subsection{X-ray luminosity and redshift sub-samples} \label{sec:Complete_Xray}
Given the varying depths of the {\em Chandra} X-ray observations over the different fields, we investigate AGN enhancement as a function of projected separation for the more complete X-ray AGN samples defined in Section~\ref{sec:AGNselection}. These X-ray AGN samples are moderate ($L_X = 10^{43.2{-}43.7}$\, erg\,s$^{-1}$ at $0.5 < z < 2$) and high $L_X$ samples ($L_X > 10^{43.7}$\, erg\,s$^{-1}$ at $0.5 < z < 3$; see Figure~\ref{fig:LX_sample}). 

\begin{figure*}
    \includegraphics[width=\textwidth]{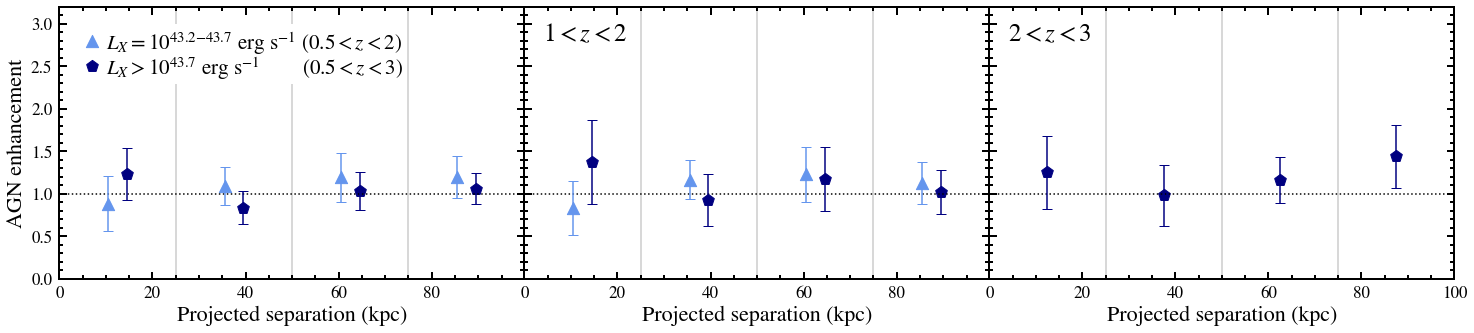}
    \caption{AGN enhancements for our varying X-ray luminosity samples (defined in Figure~\ref{fig:LX_sample}) in major galaxy pairs in bins of projected separation. \textit{Left panel}: AGN enhancement for moderate $L_X$ (sky-blue triangles) and high $L_X$ AGN (navy pentagons) over their respective complete redshifts, $0.5 < z < 2$ and $0.5 < z < 3$. \textit{Middle and right panels}: AGN enhancements in different redshift range, $1 < z < 2$ and $2 < z < 3$. The moderate $L_X$ AGN sample is not plotted in the right panel as not all fields are complete for these X-ray luminosities at $z > 2$. Vertical grey lines indicate the projected separation bin edges, and each data point is computed from all pairs in its respective bin (horizontally offset for visual aid). The data points scatter around an AGN enhancement of 1 (shown with the dotted lines), indicating that the AGN fraction in the true pair and control samples is the same across all separations.}
    \label{fig:MIDLX_HILX}
\end{figure*}

Moderate and high $L_X$ AGN enhancements for major galaxy pairs are presented in the left panel of Figure~\ref{fig:MIDLX_HILX}. Just as with the full sample of X-ray AGN, both the moderate and high $L_X$ AGN samples are consistent with no enhancement across all projected separation bins. For major galaxy pairs within $r_p < 25$\,kpc, we report enhancement values of $0.88 \pm 0.32$ and $1.23 \pm 0.30$ for moderate and high $L_X$ AGN samples, respectively. We further investigate whether these results are redshift-dependent by measuring the enhancement of moderate and high $L_X$ AGN at $1 < z < 2$ and $2 < z < 3$. As shown in the middle and right panels of Figure~\ref{fig:MIDLX_HILX}, we find no enhancement of X-ray AGN in major galaxy pairs at any redshift. \citet{Shah20} also report no enhancement across different X-ray luminosity and redshift bins. We note that with these data we are unable to rigorously investigate AGN enhancement at $z<1$ in a meaningful way due to low number statistics. While the literature provides some evidence on the redshift-dependency of the AGN-merger connection down to lower redshifts \citep[e.g.,][]{Silverman11, Goulding18}, we refrain from directly addressing this question here.

\subsection{Obscured AGN} \label{sec:Obscured_AGN}
Theoretical models predict, and observational works suggest, an increase in nuclear obscuration during advanced stages of a major merger independent of viewing angle \citep[e.g.,][]{Hopkins08, Kocevski15, Lansbury17, Ricci17, Ricci21, Blecha18, Hickox18}. As AGN emission is absorbed and re-emitted in the IR by circumnuclear dust, IR AGN selection is less susceptible to nuclear obscuration than selection via shorter wavelengths \citep[e.g.,][]{Alonso_Herrero06, Donley07, Donley10, Donley12, Hickox18}. We do not have high-quality constraints on X-ray obscuration ($N_\textrm{H}$) across all fields; furthermore, the most obscured sources could be completely undetected in the X-ray surveys. Therefore, following \citet{Satyapal14} and \citet{Weston17}, we define an ``obscured'' AGN sample as those identified exclusively in the IR. Specifically, we define this sample as those AGN which are identified in the IR (following \citealt{Donley12}) but not detected in the X-rays above the AGN threshold ($L_X > 10^{42}$\,erg\,s$^{-1}$; ``IR-only AGN''). While all X-ray-detected AGN are not necessarily unobscured, for the purposes of this analysis, we assume that the IR-only detected sample is more obscured \citep[][]{Andonie22}. Of the 709 IR AGN in our full sample, ${\sim}48\%$ are either not detected in X-rays or fall below this threshold, which corresponds to 338 ``obscured AGN.''

We calculate the fraction of ``obscured'' AGN compared to ``unobscured'' AGN in bins of projected separation, as
\begin{equation} \label{eq:obsc_frac}
    \frac{\textrm{Obscured}}{\textrm{Unobscured}} = \frac{\sum_i \mathcal{P}_\textrm{pair} \times N_{\textrm{IR-only AGN}, i}}{\sum_i \mathcal{P}_\textrm{pair} \times N_{\textrm{X-ray AGN}, i}},
\end{equation}
where, $N_{\textrm{IR-only AGN}}$ and $N_{\textrm{X-ray AGN}}$ denote the number of AGN in a pair (0, 1 or 2) selected exclusively in the IR and those selected in the X-ray (irrespective of IR selection), respectively. Just as in calculating the weighted AGN fraction, we weight our IR-only and X-ray AGN counts by their corresponding pair probabilities, and we estimate the standard error on the obscuration fraction with a common bootstrap technique \citep{Efron79, Efron81}.

\begin{figure}
    \includegraphics[width=\columnwidth]{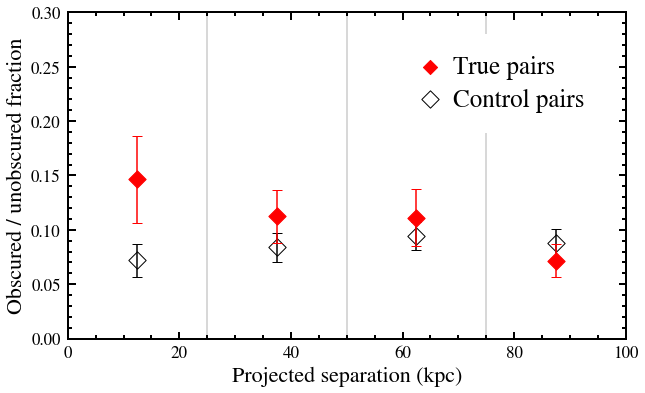}
    \caption{Obscuration fraction in bins of projected separation for major galaxy pairs. Obscuration fraction is defined as the ratio of the number of AGN selected only in the IR to the number selected in the X-ray, weighted by the appropriate pair probabilities (Equation~\ref{eq:obsc_frac}). The red and empty diamonds correspond to the obscuration fraction of true and control pairs, respectively.}
    \label{fig:OBSC_FRAC2}
\end{figure}

In Figure~\ref{fig:OBSC_FRAC2}, we plot this ratio for the true and control pair samples as a function of projected separation. For true pairs of separation $r_p > 25$\,kpc, the obscuration fraction is largely consistent with that of the control, albeit trending upward towards lower projected separations. For true pairs of separation $r_p < 25$\,kpc, however, the obscuration fraction substantially differs from that of the control pairs ($14.6 \pm 3.8\%$ versus $7.2 \pm 1.4\%$), which supports the presence of increased AGN obscuration at the smallest projected separations \citep[][see discussion in Section~\ref{sec:IR_disc}]{Satyapal14, Weston17}. 

\begin{figure} 
    \includegraphics[width=\columnwidth]{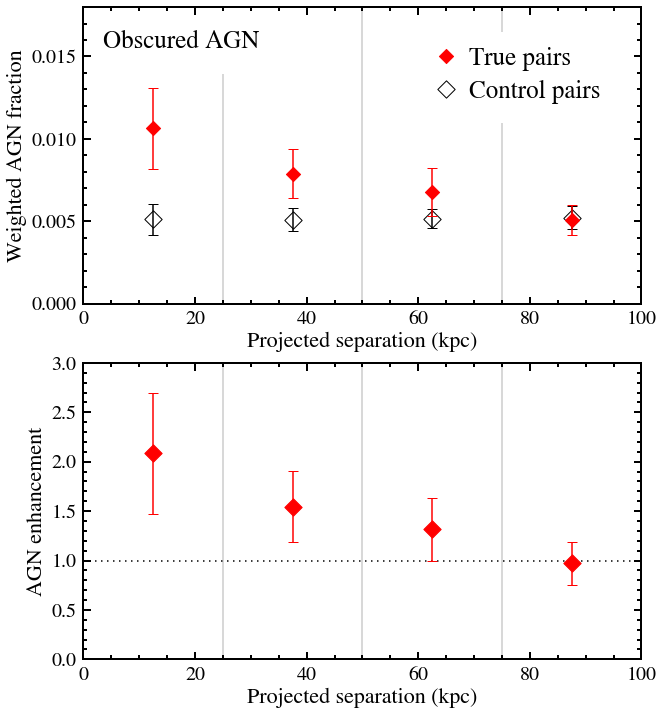}
    \caption{Weighted AGN fraction (\textit{top}) and AGN enhancement (\textit{bottom}) of obscured AGN in major galaxy pairs binned by projected separation. The red and empty diamonds correspond to the weighted AGN fractions for the true and control pairs, respectively. The AGN enhancement systematically increases from a value of 1 (dotted horizontal line; indicating no enhancement) with decreasing projected separation.}
    \label{fig:OBSC_ENH}
\end{figure}

We determine the weighted AGN fraction and AGN enhancement for our ``obscured'' AGN sample and present these results in Figure~\ref{fig:OBSC_ENH}. While the obscured AGN fraction of the {\em control group} remains stable at all separations (with $0.51 \pm 0.09$ at $r_p < 25$\,kpc), the obscured AGN fraction in true pairs increases continuously from $0.68 \pm 0.15\%$ at $50 < r_p < 75$\,kpc to $1.06 \pm 0.25\%$ at $r_p < 25$\,kpc. This corresponds to an increase in obscured AGN {\em enhancement} at decreasing projected separations peaking at $2.08 \pm 0.61$ for the closest projected pairs. However, as detailed in Section~\ref{sec:AGNfracEnh}, we can not rule out that we are slightly under-estimating the true AGN enhancement due to the challenges of dis-entangling the true and control pairs with only photometric redshift PDFs.

\section{Discussion} \label{sec:Discussion}
Due to our new pair probability approach, we have been able to incorporate the largest (known) sample of high-redshift galaxy pairs into a merger-induced AGN enhancement calculation{\footnote{To give an indicative number of the sample size, we report that 22,295 major galaxy pairs of projected separation $r_p < 100$\,kpc have a true pair probability $\mathcal{P}_\textrm{pair} > 0.01$ (see Table~\ref{tab:all_XAGN}).}. We begin this section by discussing the advantages of our pair probability method compared to other pair-finding techniques that make use of photometric redshift PDFs (Section~\ref{sec:Disc-PDF}). Then, we discuss our findings on the connection between close galaxy pairs and AGN enhancement in the context of previous work and a merger-driven evolutionary scenario for heavily dust-obscured black hole growth (Section~\ref{sec:AGN_merger_connection}).

\subsection{Photometric PDF approaches for pair probabilities} \label{sec:Disc-PDF}
Our method for calculating the probability that two galaxies are in a physically associated pair involves calculating the probability that two galaxies are within a relative line-of-sight velocity of $|\Delta V| < 1000$\,km\,s$^{-1}$ (Section~\ref{sec:P_DV}) and a projected separation of $r_p < 100$\,kpc (Section~\ref{sec:P_r}) from a convolution of redshift photometric redshift PDFs or spectroscopic redshifts (Section~\ref{sec:P_pair}). A major strength of our probabilistic approach is that we include the full underlying uncertainties of the photometric redshifts, but we neither need to exclude photometric redshifts based on some arbitrary uncertainty threshold nor apply a correction for chance (not physically associated) projected pairs, which breaks down for dense environments \citep[e.g.,][]{LeFerve00, Kartaltepe07, Bundy09}. 

Another method has previously been developed to circumvent the challenges of using photometric redshift PDFs with a broad range of quality/widths \citep[][]{LS15}. This method was expanded by \citet{Mundy17} and \citet{Duncan19} to calculate weighted merger fractions in the CANDELS fields while correcting for stellar mass completeness. For galaxy 1 and 2 in a projected pair, the method compares photometric redshift PDFs, $P(z)$, using the combined redshift probability function, $\mathcal{Z}(z)$, defined as:
\begin{equation} \label{eq:combined_z}
    \mathcal{Z}(z) = \frac{2 \times P_1(z) \times P_2(z)}{P_1(z) + P_2(z)}.
\end{equation}
A numerical pair probability that these galaxies are physically associated, $\mathcal{P}_\textrm{LS15}$, is then obtained by taking the cumulative integral of $\mathcal{Z}(z)$ over the full redshift range:
\begin{equation} \label{eq:LS_prob}
    \mathcal{P}_\textrm{LS15} = \int_0^z \mathcal{Z}(z)dz.
\end{equation}
 
We show the resulting integrals and ${P}_\textrm{LS15}$ values, for the example PDFs in Figure~\ref{fig:conv_method} (see the black curves and inset text in the middle-left and bottom-left panels). This approach effectively determines pair probability based on the amount of {\em overlap} of the two photometric redshift PDFs. For a set of narrow PDFs, the method is successful in establishing which of the galaxies are most likely to be physically associated. For example, in the middle-left panel of Figure~\ref{fig:conv_method}, the PDFs are well constrained and their peak values are close. The combined redshift method gives a pair probability of $\mathcal{P}_\textrm{LS15} = 0.69$. However, the PDFs of the projected pair in the bottom left panel are very poorly constrained, i.e., they are very broad and feature multiple peaks. In this case, Equation~\ref{eq:LS_prob} gives a probability of $\mathcal{P}_\textrm{LS15} = 0.75$, which is larger than the previous example for the well-defined PDFs. The values of our relative line-of-sight probabilities for the same example pairs are $\mathcal{P}_{\Delta V} = 0.33$ and $\mathcal{P}_{\Delta V} = 0.05$, respectively (for the middle and bottom row examples in Figure~\ref{fig:conv_method}). These $\mathcal{P}_{\Delta V}$ values appear to more clearly reflect the relative quality of the underling PDFs, and hence the greater uncertainty in the two galaxies being in physically associated pairs. 
 
In contrast to assessing the overlap between two PDFs, our convolution method results in pair probabilities that are larger when the peaks of the two PDFs are similar but smaller when the individual PDF widths are larger. This guarantees that sources with less certain redshifts will always be assigned a smaller probability of being in a physically associated pair. 

Whilst a ``brute-force'' Monte-Carlo approach can achieve the same results as the convolution method (i.e., by randomly drawing redshifts from PDFs and calculating $\Delta V$ for all projected pairs each iteration; see Figure~\ref{fig:conv_method}), the convolution approach is significantly more efficient due to the computational ease of calculating a convolution as the product of two fast Fourier transforms. This makes it applicable to the large samples required to rigorously investigate AGN activity, or any other galaxy properties, in galaxy pairs as a function of projected separation.

\subsection{The association of galaxy pairs with different AGN types} \label{sec:AGN_merger_connection}
Studies of low-redshift galaxies find an excess of disturbed morphological features indicative of a galaxy merger or interaction in AGN, compared to samples of inactive galaxies \citep[e.g.,][]{Guyon06, RamosAlmeida11, Goulding18, Koss18, Ellison19, Gao20, Pierce22,Pierce23}. An enhancement of AGN is also found in physically associated projected pairs, relative to matched control galaxies, but with the quantitative results being dependent on the AGN selection method \citep[e.g.,][]{Ellison11, Ellison13b, Satyapal14}. In this work we have investigated the connection between physically associated galaxy pairs and the number of X-ray and IR-selected AGN at cosmic noon ($0.5 < z < 3$). We have built on the work of \cite{Shah20}, who took a similar approach but was limited to using only spectroscopic redshifts to identify which galaxies are in physically associated pairs. In the following, we discuss our results for X-ray AGN and the most obscured AGN in the context of an AGN-merger connection at cosmic noon.

\subsubsection{X-ray AGN are not enhanced in close pairs at cosmic noon} \label{sec:Xray_disc}
Our results are consistent with no enhancement for X-ray-selected AGN (i.e., $L_X > 10^{42}$\,erg\,s$^{-1}$) in major (or minor) galaxy pairs at close projected separations (see Figure~\ref{fig:XAGN_IRAGN}). This is strong evidence that X-ray AGN are not preferentially associated with galaxy mergers/interactions in the CANDELS and COSMOS fields. This is in agreement with the spectroscopic redshift sample of \cite{Shah20}. Many other observational studies at high redshift also report no merger-induced AGN enhancement in X-ray AGN populations. Working in the COSMOS field, \citet{Cisternas11} find no difference in the frequency of morphological disturbances indicative of an ongoing merger between X-ray-selected AGN and inactive galaxies at $0.3 < z < 1$. Likewise, \citet{Kocevski12} report that X-ray AGN and inactive control galaxies at $1.5 <z < 2.5$ are highly disturbed at consistent rates in GOODS-S ($16.7^{{+}5.3}_{{-}3.5}\%$ versus $15.5^{{+}2.8}_{{-}2.2}\%$). Both these works note that the majority of X-ray AGN reside in disk galaxies, suggesting secular processes to be the more efficient fueling mechanism for these systems \citep[see also,][]{Rosario15}. \cite{Silva21} also find no difference in the fraction of multi-wavelength selected AGN in merger and non-merger galaxies at cosmic noon in the CANDELS fields.

We also find no difference in enhancement between our moderate ($L_X = 10^{43.2{-}43.7}$\,erg\,s$^{-1}$) and high $L_X$ (${>}10^{43.7}$\,erg\,s$^{-1}$) AGN samples in major galaxy pairs (see Figure~\ref{fig:MIDLX_HILX}). Specifically, for both samples the AGN enhancement in galaxy pairs separated by less than 25\,kpc, is consistent with 1 (i.e., no enhancement; $0.88 \pm 0.32$ and $1.23 \pm 0.30$, respectively). No trend of AGN enhancement with X-ray luminosity is similarly reported by other works at cosmic noon \citep[e.g.,][]{Hewlett17, Villforth14, Villforth17, Shah20}.

Our results are not unexpected when compared to recent cosmological simulations. AGN frequency in merging galaxies has been investigated in \texttt{EAGLE} \citep{McAlpine20}, which identifies AGN based on a bolometric luminosity threshold ($L_\textrm{bol} > 10^{43}$\,erg\,s$^{-1}$; also see \citealt{ByrneMamahit22} for similar results based on \texttt{IllustrisTNG}). If we assume that X-ray emission is a reasonable tracer of instantaneous accretion rates, we can make an apt comparison to this work. For the nearby Universe ($0.05 < z < 0.1$), \cite{McAlpine20} report a modest AGN enhancement (${\sim}1.3$) at physical separations less than 30\,kpc. However, at $z > 1$, they report no enhancement even for the most rapidly accreting SMBHs ($\lambda_\textrm{edd} > 0.01$), which is consistent with our analysis of X-ray selected AGN in galaxy pairs at $0.5 < z < 3$.

All together, the evidence supports that X-ray AGN are not preferentially associated with galaxy interactions or mergers at cosmic noon. However, it is critical to consider AGN selection. For example, \citet{Secrest20} find no statistically significant enhancement (a value of $2.22^{{+}4.44}_{{-}2.22}$) of X-ray AGN in post-mergers at $0.03 < z < 0.15$, which is in stark contrast to the ${\sim}17$ factor excess of MIR AGN they find in the same sample. Furthermore, \citet{Satyapal14} report a ${\sim}4{-}6{\times}$ excess of IR-selected AGN in spectroscopic galaxy pairs at $0.01 < z < 0.2$ separated by less than 10\,kpc, while \citet{Ellison13b} find only a ${\sim}2.5{\times}$ excess in AGN selected from optical emission lines at the same redshift. Using identical MIR and optical AGN selection, \citet{Ellison19} report that almost $60\%$ of MIR AGN are visually disturbed (indicative of a post-merger), double that of optically identified AGN \citep[see also,][]{Bickley23}. Therefore, in the following we discuss the results for the most ``obscured AGN'' at cosmic noon. 

\subsubsection{Obscured AGN enhancement in close pairs at cosmic noon} \label{sec:IR_disc}
We find evidence for enhanced AGN obscuration at late merging stages (${<}25$\,kpc), defined by an increased fraction of AGN identified only through their IR emission and not through X-rays (Figure~\ref{fig:OBSC_FRAC2}). 
Additionally, we observe evidence for an increasing AGN enhancement at decreasing projected separations in this obscured AGN sample (Figure~\ref{fig:OBSC_ENH}). Our results qualitatively match results at low redshift, which together might favour merger-induced obscuration and enhanced black hole growth. For example, \citet{Weston17} report a similar excess in obscuration fraction (factor of ${\sim}2{-}6$) when comparing the MIR- to optical emission-line-selected AGN in merging and isolated nearby galaxies. \citet{Satyapal14} find that the ratio of AGN selected exclusively in the MIR (from \textit{WISE} colour cuts) to that selected in the optical (i.e., emission-line Seyferts; may also be selected in the MIR), increases significantly for pairs separated by less than 30\,kpc compared to a matched control sample. They report ratio excesses of ${>}2$ and ${\sim}4$ for pairs separated by less than 10\,kpc and for visually identified post-mergers, respectively. 

IR AGN selection is more effective during the most obscured phases of black hole growth, compared to optical emission-line or X-ray selection, which can be faint or completely invisible \citep[e.g.,][]{Sanders88a, Sanders88b, Veilleux09,GouldingAlexander09,Snyder13, Kocevski15, Blecha18, Hickox18, Andonie22}. It follows that if obscured galaxy growth is linked to merging and interacting galaxies, then AGN selection most sensitive to this population would exhibit the strongest AGN enhancement. Indeed, simulations predict that intense nuclear starbursts induced during the late stage of a gas-rich major merger obscure the SMBH with hot dust and gas \citep{Cattaneo05, Hopkins08} and that the SMBH may accrete significant mass during a relatively short obscured phase \citep[e.g.,][]{Fabian99, Treister09, DraperBallantyne10}. Whilst mergers may not be relevant for the majority of AGN or black hole growth \citep[e.g.,][]{McAlpine20,ByrneMamahit22}, these discrete events may play an important evolutionary phase for the host galaxy and host halo, due to boosted AGN luminosities, disruptions of the galaxy disk and consequently effective AGN feedback \citep[][]{Quai23,Davies23}.

Robust and uniformly measured X-ray column densities are not available for our X-ray AGN sample, which could provide another method to investigate trends with obscuration. However, we note that at low redshift, \citet{Koss18} report a significant enhancement of late-stage nuclear mergers in AGN selected in {\em hard} X-rays ($14{-}195$\,keV from \textit{Swift}/BAT), suggesting soft X-rays (i.e., those observed by {\em Chandra}) are unable to detect AGN in the heavily obscured phase of a merger. Furthermore, for cosmic noon AGN, using an X-ray spectral analysis of 154 $z \sim 1$ AGN in the CANDELS fields, \citet{Kocevski15} find that the frequency of disturbed morphologies, indicative of a merger/interaction, in Compton-thick AGN is nearly three times that of an unobscured AGN sample ($21.5^{{+}4.2}_{{-}3.3}\%$ versus $7.8^{{+}1.9}_{{-}1.3}\%$). However, many obscured AGN - and the ones most associated with mergers based on these results - may be hidden from the {\em Chandra} X-ray surveys covering the CANDELS and COSMOS fields (\citealt{Andonie22}). Indeed, this is strongly supported by \citet{Donley18}, who find in the CANDELS portion of the COSMOS field that AGN detected only in the IR (i.e., receptive to obscured AGN) are more than twice as likely than those selected exclusively in the X-ray ($75^{{+}8}_{{-}13}\%$ versus $31^{{+}6}_{{-}6}\%$) to have disturbed host morphologies indicative of a merger/interaction. Therefore, merger-induced obscuration could explain why an enhancement of AGN is not seen for the typical X-ray selected samples at cosmic noon. 

We are in agreement with several other works that there seems to be an increased association between obscured AGN and interactions/mergers, compared to unobscured AGN. However, it is important to address the differences and limitations among different AGN selection methods. For example, large samples of IR AGN are often selected at low redshift via a colour-colour cut. Such a selection requires AGN emission to dominate over that of the host \citep[][]{Mateos12, Gurkan14}. Beyond $z \gtrsim 1$, strong stellar emission compounds this luminosity bias, making selecting clean samples of IR AGN difficult \citep{Blecha18}. Stellar contamination is addressed by \citet{Donley12}, who implement a power law selection, in addition to a colour-colour cut, to identify clean AGN samples from \textit{Spitzer}/IRAC data out to $z \sim 3$. Such a strict criteria again requires dominant AGN emission in the IR, making this selection conservative towards the most luminous AGN. Therefore, more complete and confident samples of IR AGN are required to better constrain their intrinsic properties, their connection to interactions/mergers and a robust comparison to luminosity-matched samples of unobscured AGN. This will be possible with very large samples that are covered by comprehensive spectral energy distribution analyses, with improved constraints on X-ray column densities and with improved quality infrared photometry from {\em JWST} \citep[e.g.,][]{Satyapal21,Andonie22} . 

Finally, we note that our results on the most obscured sources would not be possible without the inclusion of photometric redshifts because the obscured AGN sample has significantly less spectroscopic coverage ($8.3\%$) than X-ray AGN ($30\%$). Future spectroscopic surveys, targeting the most obscured AGN, will provide further insight into this important population \citep[e.g.,][]{Merloni19}.

\section{Conclusions} \label{sec:Conclusion}
We have developed a novel technique for identifying high probability projected galaxy pairs (i.e., those highly likely to be physically interacting with line-of-sight velocities within $\pm$1000\,km\,s$^{-1}$) using convolutions of photometric redshift probability distribution functions. This enables us to compile the largest sample of high-redshift galaxy pairs ever used for a merger-induced AGN enhancement calculation. Our work is an important addition to studies limited to only spectroscopic redshifts, as they lack complete information on the line-of-sight velocity for {\em all} galaxy pairs within a given projected separation. 

We use our technique to translate pair probabilities into weights when calculating the fraction of X-ray and IR AGN in $0.5 < z < 3$ projected major ($M_{*,1} > 10^{10} M_\odot$ and $M_{*,1}{:}M_{*,2} < 4{:}1$) and minor galaxy pairs ($M_{*,1} > 10^{10.4} M_\odot$ and $4{:}1 < M_{*,1}{:}M_{*,2} < 10{:}1$) in the CANDELS and COSMOS fields. We take a complementary probabilistic approach to identify highly likely isolated galaxies from which we assemble a control sample matched to the true pairs in redshift, stellar mass, environment and redshift quality. We compute the AGN enhancement as the ratio of the weighted AGN fractions of the true pairs to the corresponding control group. Our key findings are:
\begin{enumerate}
    \item We find no evidence of an enhancement of X-ray AGN ($L_{X}>10^{42}$\,erg\,s$^{-1}$) in major or minor galaxy pairs for projected separations of 5--100\,kpc. Specifically, in the closest separation bin of 5--25\,kpc we find an AGN enhancement values of $0.98 \pm 0.19$ and $1.02 \pm 0.15$ for the minor and major pairs, respectively (Figure~\ref{fig:XAGN_IRAGN}).
    \item We further explore two X-ray AGN samples more complete to the sensitivity of observations across all fields investigated: (1) moderate luminosity ($L_X{=}10^{43.2{-}43.7}$\,erg\,s$^{-1}$; $z = 0.5{-}2$) and; (2) high luminosity ($L_X{>}10^{43.7}$\,erg\,s$^{-1}$; $z = 0.5{-}3$). We still find no evidence for an AGN enhancement in the close galaxy pairs, nor do we observe differences as a function of redshift or X-ray luminosity 
    (Figure~\ref{fig:MIDLX_HILX}).
    \item Defining the most ``obscured AGN'' as those that are identified in the IR \citep[][]{Donley12} but are not detected in the X-rays, we find the fraction of obscured AGN increases with decreasing projected separation. 
    Specifically, we find that major galaxy pairs separated by less than 25\,kpc are twice as likely to be obscured compared to isolated controls (Figure~\ref{fig:OBSC_FRAC2}). 
    \item We find a systematic trend of increasing AGN enhancement as a function of decreasing projected separation for these ``obscured AGN''. This begins at a projected separation of ${\sim}75$\,kpc and peaks with an AGN enhancement value of $2.08\pm0.61$ for separations of 5--25\,kpc (Figure~\ref{fig:OBSC_ENH}).
\end{enumerate}
Our results for X-ray selected AGN are consistent with previous work of galaxy pairs at cosmic noon that were limited to only using spectroscopic redshifts \citep[][]{Shah20}. This indicates no evidence for an association between galaxy interactions/mergers and X-ray selected AGN. In contrast, our results hint towards a potentially strong association between galaxy interactions and the most obscured AGN, which population cannot currently be adequately investigated when limited only to samples with spectroscopic redshifts. These results are broadly consistent with an evolutionary scenario of merger-driven obscured black hole growth. 

This work is limited by the quality of infrared photometry and limited spectroscopic redshift confirmation for the most obscured AGN. For example, we cannot rule out that we are underestimating the true AGN enhancement due to the challenges of disentangling the true and control pairs with only photometric redshift PDFs. {\em JWST} will provide higher quality imaging in the near- and mid-IR, improving source photometry, in particular for close galaxy pairs, and placing higher quality constraints on the spectral energy distributions of AGN \citep[e.g.,][]{Satyapal21, Rigby23}. Furthermore, future near-IR spectroscopic surveys (e.g., MOONS; \citealt{MOONS} and PFS; \citealt{PFS}) will provide increasingly higher spectroscopic completeness of deep fields, permitting pair identifications with much higher certainty than possible with photometric redshifts. Therefore, the coming years will yield ground-breaking insight into the drivers of black hole growth at cosmic noon, in particular for the most obscured systems.

\section*{Acknowledgements}
We thank the referee, John Silverman, for constructive comments. CMH acknowledges funding from an United Kingdom Research and Innovation grant (code: MR/V022830/1). We thank Danielle Leonard and Anne Archibald for constructive discussions and suggestions. During this work we made use of the {\tt Python} packages {\tt matplotlib} (\citealt{matplotlib}),  {\tt astropy} (\citealt{astropy1,astropy2}), {\tt numpy} (\citealt{numpy}), {\tt pandas} (\citealt{pandas2, pandas1}) and {\tt scipy} (\citealt{scipy}).

\section*{Data Availability}
All photometric data, redshift information and derived quantities (e.g., stellar masses) used in this work are available in public catalogues as described in Section~\ref{sec:Method}. The derived values of this work, presented in Figures \ref{fig:XAGN_IRAGN}--\ref{fig:OBSC_ENH}, are tabulated in Appendix~\ref{app:tables}. Code used in this work is available on github (\href{https://github.com/sldough21/pair-convolutions}{https://github.com/sldough21/pair-convolutions}).

\bibliographystyle{mnras}


\appendix

\section{Spectroscopic redshifts and comparison to photometric redshifts}
\label{App:Spectroscopy}
\label{app:phot_val}

Here we provide the references that were used to compile the spectroscopic redshifts described in Section~\ref{sec:Spectroscopy}, followed by a comparison between the spectroscopic redshift and photometric redshifts described in Section~\ref{Sec:PDFs}. Table~\ref{Table:redshifts} summarises the total number of galaxies (parent sample and AGN), the fraction of these with spectroscopic redshifts and the accuracy of the corresponding photometric redshift PDFs ($\sigma_\textrm{NMAD}$).

For the CANDELS GOODS-N field, we utilised the spectroscopic redshifts that were obtain on: (1) Keck I, with the Low Resolution Imaging Spectrometer \citep[LRIS;][]{Cohen00, Cohen01, Dawson01, Cowie04, Reddy06} and the Multi-Object Spectrometer For Infra-Red Exploration \citep[MOSFIRE;][]{Kriek15, Wirth15}; (2) Keck II, with the DEep Imaging Multi-Object Spectrograph \citep[DEIMOS;][]{Cowie04, Wirth04, Barger08, Cooper11}; and (3) on Subaru with the Multi-Object Infrared Camera and Spectrograph \citep[MOIRCS;][]{Barger08}. 

For the CANDELS GOODS-S field, we used spectroscopic redshifts obtained on: (1) Keck I using MOSFIRE \citep[][]{Kriek15}; (2) Keck II with DEIMOS \citep[][]{Silverman10, Cooper12}; and (3) the Very Large telescope (VLT) with the VIsible MultiObject Spectrograph \citep[VLT/VIMOS;][]{Ravikumar07, Balestra10, LeFevre13, Garilli21}, the FOcal Reducer and low dispersion Spectrograph 1 and 2 \citep[FORS1/2;][]{Vanzella08, Kurk13}, and the Multi Unit Spectroscopic Explorer \citep[MUSE;][]{Inami17, Urrutia19}.

The spectroscopic redshifts used in this study for the CANDELS EGS field were obtained on Keck I with LRIS and MOSFIRE \citep[][]{Kriek15, Masters19} as well as on Keck II with DEIMOS \citep[][]{Cooper12a, Matthews13, Masters19}.

For the CANDELS UDS field, the data were taken on: (1) Keck I with LRIS and MOSFIRE \citep[][]{Kriek15, Masters19}; (2) Keck II with DEIMOS \citep[][]{Masters19}; (3) VLT with FORS2 \citep[][]{Bradshaw13} and VIMOS \citep[][]{Bradshaw13, Scodeggio18, Garilli21}; and (4) Subaru with the Faint Object Camera And Spectrograph \citep[FOCUS;][]{Yamada05}.

Finally, for all of COSMOS, spectroscopic observations were taken on: (1) Keck I with LRIS and MOSFIRE \citep[][]{Kriek15, Nanayakkara16, Casey17, Masters19, Stanford21}; (2) Keck II with DEIMOS \citep[][]{Capak04, Kartaltepe10, Hasinger18, Masters19, Stanford21}; (3) VLT with FORS2 \citep[][]{Comparat15}, VIMOS \citep[][]{Lilly07, Straatman18, vanderWel21} and the $K$-band Multi-Object Spectrograph \citep[KMOS;][]{Euclid20}; (4) Subaru with the Fiber Multi-Object Spectrograph \citep[FMOS;][]{Kartaltepe15, Silverman15} and MOIRCS \citep[][]{Onodera12}; and (5) the Magellan Telescope with the Inamori-Magellan Areal Camera and Spectrograph \citep[IMACS;][]{Trump07, Trump09}. 

\begin{table}
	\centering
	\caption{The number of galaxies (N) in the parent sample and the number of AGN split into CANDELS and COSMOS. $f_{\rm spec}$ is the percentage of sources in each category with a spectroscopic redshift and $\sigma_{\rm NMAD}$ is the normalized median absolute deviation between $z_{\rm spec}$ and $z_{\rm phot}$ (see Equation~\ref{eq:NMAD}).}
	\label{tab:PDF_point_statstics}
	\begin{tabular}{l c c c} 
            \hline \hline
		~ & N & $f_\textrm{spec}$ (\%) & $\sigma_\textrm{NMAD}$   \\ \hline
            Parent Sample (CANDELS) & 22093 & 19.9 & 0.022 \\
            AGN (CANDELS) & 1036 & 34.1 & 0.033 \\
            Parent Sample (COSMOS) & 112436 & 5.3 & 0.015 \\
            AGN (COSMOS) & 2216 & 24.3 & 0.033 \\
\hline
	\end{tabular}
 \label{Table:redshifts}
\end{table}

We are primarily interested in the fraction of AGN in paired galaxies identified via their photometric redshift PDFs. Therefore, it is important to have an assessment of the quality of the photometric redshifts for AGN, compared to the wider galaxy population. In Figure~\ref{fig:ZPHOT_val} we compare spectroscopic redshifts ($z_{\rm spec}$) to the peak of the PDF estimate of the photometric redshift ($z_{\rm phot}$). We reiterate that these $z_{\rm phot}$ values are not used in the analyses to identify true galaxy pairs, but instead we use the full PDFs to encapsulate the full uncertainties. Therefore, the comparison of $z_{\rm spec}$ and $z_{\rm phot}$ values can only be considered indicative of the quality of the PDFs and thus hides the full treatment of the uncertainties. 

\begin{figure}
    \includegraphics[width=\columnwidth]{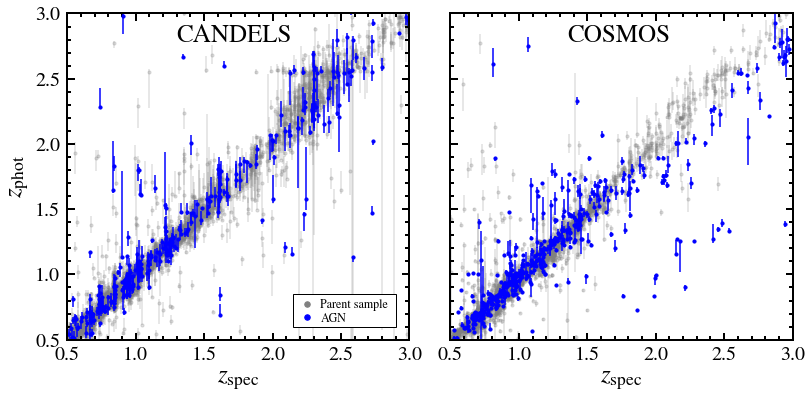}
    \caption{Spectroscopic redshifts ($z_\textrm{spec}$) versus corresponding photometric redshfit PDF peak estimate ($z_{\textrm{phot}}$) for those done by the CANDELS \citep[\textit{left};][]{Kodra22} and COSMOS teams \citep[\textit{right};][]{Weaver22}. The grey and blue points correspond to the parent sample and all AGN,  respectively. Error bars correspond to the upper and lower $68\%$ confidence intervals of the photometric redshift PDFs.}
    \label{fig:ZPHOT_val}
\end{figure}

Figure~\ref{fig:ZPHOT_val} shows that for the majority of the parent sample (grey data points) and the AGN (blue data points) the two redshifts are in good agreement. One way to quantify this is with the normalized median absolute deviation \citep[NMAD;][]{Hoaglin83} comparing the spectroscopic and photometric redshifts, which is a common approach for photometric redshift validation \citep[e.g.,][]{Guo13, Nayyeri17, Stefanon17, Barro19, Weaver22}. This is defined as
\begin{equation} \label{eq:NMAD}
    \sigma_\textrm{NMAD} = 1.48 \times \textrm{median}\left(\frac{|\Delta z|}{1+z_\textrm{spec}}\right),
\end{equation}
where $\Delta z$ is the difference between the peak photometric redshift PDF estimate and the corresponding spectroscopic redshift. 

The CANDELS and COSMOS PDFs are exceptionally accurate for both the parent sample ($\sigma_\textrm{NMAD} \sim 0.02$) and the AGN ($\sigma_\textrm{NMAD} \sim 0.03$), albeit slightly poorer for the AGN. The AGN also have higher incidence of spectroscopic redshifts compared to the parent sample (34\% compared to 20\% for CANDELS and 24\% compared to 5\% for COSMOS). This will further improve the quality of true pair probabilities presented in Section~\ref{sec:P_pair}. Overall, we feel confident the photometric PDFs are of sufficiently high quality for our large statistical study, and future observations with improved infrared photometry and larger spectroscopic samples will provide even tighter constraints (see discussion in Section~\ref{sec:Conclusion}).

\section{Probabilistic approach to selecting control galaxies}\label{app:control_selection}
Here we provide a comprehensive overview of the matching algorithm used to select control galaxy pairs in Section~\ref{sec:Control}. 

As we calculate the weighted AGN fraction for galaxy pairs (Section~\ref{sec:AGNfracEnh}), we assemble isolated galaxy ``pairs'' as our control sample. In other words, we match physically associated pairs, with weights $\mathcal{P}_\textrm{pair}$, to physically unassociated pairs. By construction, we require unassociated pairs to have $\mathcal{P}_\textrm{pair} = 0$, which requires either $\mathcal{P}_{\Delta V} = 0$ or $\mathcal{P}_r^\textrm{0-100 kpc} = 0$ (Equation~\ref{eq:trueP}). In Section~\ref{sec:P_DV}, we described how $\mathcal{P}_{\Delta V}$ is derived from the $P(z)$ of paired galaxies independent of projected separation. Therefore, we select unassociated pairs on the basis of $\mathcal{P}_r^\textrm{0-100 kpc} = 0$, and we interpret their $\mathcal{P}_{\Delta V}$ as weights both statistically comparable to those of the true pair sample and reflective of the individual unassociated galaxy redshift uncertainties.

For each true pair (TP), we select 3 unassociated control pairs (CP) matched in redshift, mass, environmental density, $P(z)$ width\footnote{For matching purposes, sources with secure spectroscopic redshifts are assigned a $P(z)$ width equal to 0.01.} and relative line-of-sight probability from a pool of galaxies highly likely to be physically isolated. We begin by assuming that every galaxy in our parameter space has some probability of being isolated (i.e., no companion within $r_p < 100$ kpc or $|\Delta V| < 1000$ km\,s$^{-1}$), $\mathcal{P}_{\textrm{iso}}$, defined as
\begin{equation}
    \mathcal{P}_{\textrm{iso}} = \prod_{i=1}^{N} (1 - \mathcal{P}_{\textrm{pair},i}^\textrm{5-100 kpc}),
\end{equation}
where $N$ is the total number of its projected pairs and $\mathcal{P}_{\textrm{pair},i}^\textrm{5-100 kpc}$ is the true pair probability (Equation~\ref{eq:trueP}) of pair $i$ evaluated over the full separation range. We create a pool of all galaxies with $\mathcal{P}_{\textrm{iso}} > 90\%$, from which we will select controls. 

Following \citet{Ellison13b}, we compute the local environmental density for each galaxy, $\Sigma$, defined as
\begin{equation} \label{eq:envdens}
    \Sigma = \frac{n}{\pi r_n^2},
\end{equation}
where $r_n$ is the projected separation to the $n$th nearest companion within $|\Delta V| < 1000$ km\,s$^{-1}$. We set $r_n = 1$ Mpc and apply our probabalistic methodology to measuring $n$ as the sum of all true pair probabilities of all projected companions within 1 Mpc:
\begin{equation} \label{eq:envdens2}
    n = \sum_{i=1}^N \mathcal{P}_{\textrm{pair,i}}^\textrm{0-1 Mpc}.
\end{equation}

In finding the best matches, we would ideally calculate the difference in mass, redshift, environment, $P(z)$ width (for both galaxies) as well as $\mathcal{P}_{\Delta V}$ between the TP and all candidate CPs. This approach would involve first calculating $\mathcal{P}_{\Delta V}$ then sorting for over 6 billion unique galaxy pairs. Rather, we construct an iterative search algorithm to find sufficient matches in a timely manner. For both galaxies in a TP, we assemble initial samples of control candidates with $|\Delta z| < 0.05$, $|\Delta \textrm{log}(M_*)| < 0.05$, $|\Delta \Sigma| < 2$ and $|\Delta \textrm{log}(P(z)\ \textrm{width})| < 0.05$. We generate every combination of galaxies from the two samples separated by more than 20\,arcsec to ensure $\mathcal{P}_r^\textrm{0-100 kpc}=0$, and we compute $\mathcal{P}_{\Delta V}$ for each pairing. In order for a pair of relative line-of-sight probability, $\mathcal{P}_{\Delta V}^{\textrm{CP}}$, to be selected as a match to a true pair of relative line-of-sight probability, $\mathcal{P}_{\Delta V}^{\textrm{TP}}$, it must meet the criteria of our matching function:
\begin{equation} \label{eq:matching_function}
    |\textrm{log} \left( \mathcal{P}_{\Delta V}^{\textrm{TP}} \right) - \textrm{log} \left( \mathcal{P}_{\Delta V}^{\textrm{CP}} \right)| < \frac{1.54 \times 10^{-4}}{\mathcal{P}_{\Delta V}^{\textrm{TP}} + 5.24 \times 10^{-5}} + 0.05.
\end{equation}
We devised the matching function such that high probability TPs (i.e., the ones that receive the highest weights in the weighted AGN fraction calculation), are strictly matched in $\mathcal{P}_{\Delta V}$ while the matching to low probability TPs (i.e., those whose weights will contribute negligibly to the weighted AGN fraction), is less stringent. Matched $\mathcal{P}_{\Delta V}$ are shown in the bounds of the matching function in Figure~\ref{fig:PdV_match}. 

\begin{figure}
    \includegraphics[width=\columnwidth]{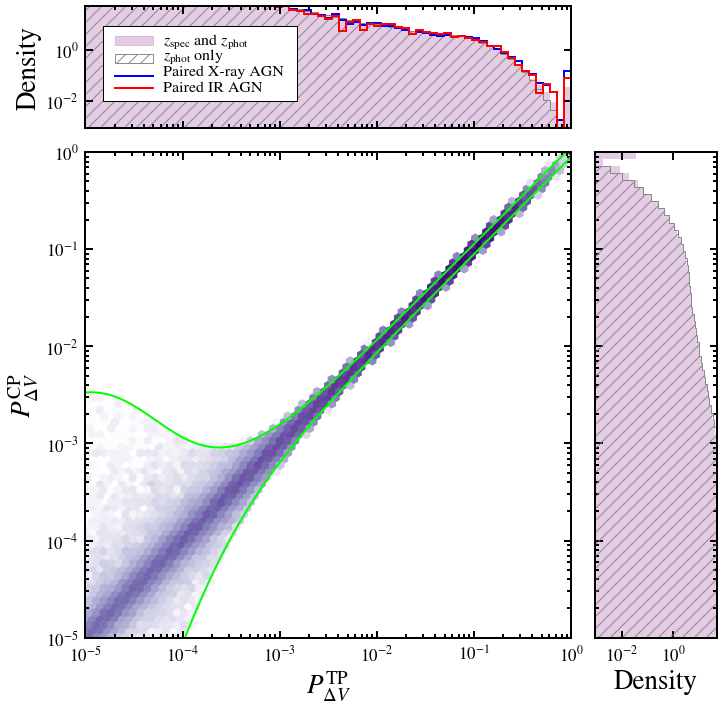}
    \caption{Distribution of the relative line-of-sight probabilities of the control pairs, $\mathcal{P}_{\Delta V}^\textrm{CP}$, matched to that of the true pairs, $\mathcal{P}_{\Delta V}^\textrm{TP}$. The green lines correspond to the matching function extrema (Equation~\ref{eq:matching_function}) projected onto the $\mathcal{P}_{\Delta V}^\textrm{TP}{-}\mathcal{P}_{\Delta V}^\textrm{CP}$ plane. True pairs with high $\mathcal{P}_{\Delta V}^\textrm{TP}$ are tightly matched by $\mathcal{P}_{\Delta V}^\textrm{CP}$ to ensure equal weighting when calculating, then comparing, weighted AGN fractions (Equation~\ref{eq:weightedAGNfrac}). Distributions of $\mathcal{P}_{\Delta V}^\textrm{TP}$ and $\mathcal{P}_{\Delta V}^\textrm{CP}$ are shown with (solid purple histogram) and without (grey striped histogram) the inclusion of spectroscopic redshifts, adjacent to the corresponding axes. Normalised distributions of $\mathcal{P}_{\Delta V}^\textrm{TP}$ for pairs featuring X-ray and IR AGN are given by blue and red lines, respectively}
    \label{fig:PdV_match}
\end{figure}

If more than 3 candidate CPs satisfy the matching function, then we select the 3 best-matched CPs by minimizing the difference:
\begin{equation}
\begin{aligned}
    \textrm{dif.} &= (\Delta z_1)^2 + (\Delta \textrm{log}(M_{*,1}))^2 + \frac{(\Delta \Sigma_{1})^2}{10} + (\Delta \textrm{log}(P_1(z)\ \textrm{width}))^2 \\
    & + (\Delta z_2)^2 + (\Delta \textrm{log}(M_{*,2}))^2 + \frac{(\Delta \Sigma_{2})^2}{10} + (\Delta \textrm{log}(P_2(z)\ \textrm{width}))^2 \\ 
    & + 10(\Delta \textrm{log}(\mathcal{P}_{\Delta V})^2,
\end{aligned}
\end{equation}
where the subscripts 1 and 2 indicate the difference in matched parameters to galaxy 1 and 2 in a TP. We amplify the difference in $\mathcal{P}_{\Delta V}$ so that the CPs receive the closest weight possible to their matched TP. Additionally, we down-weight the environmental components to place their difference in roughly the same order of magnitude as that of the other matched parameters (i.e., stellar masses range from $\textrm{log}\left(M_*/M_\odot\right) \sim 9.4{-}12$ while environmental density ranges from $\Sigma \sim 0{-}25$\,Mpc$^{-1}$; see Figure~\ref{fig:control_match}). If less than 3 CPs satisfy the matching function, then we iteratively broaden the initial $z$, log($M_*$), $\Sigma$ and log($P(z)$ width) search thresholds by 0.03, 0.03, 1 and 0.05, respectively, until 3 do. CPs are selected with replacement for different TPs, and we require 6 unique galaxies to comprise the 3 CPs per TP to mitigate duplicate effects. Normalized distributions of the matched parameters with and without weights are shown in Figure~\ref{fig:control_match}, which reveals the matching process was successful.

\section{Simulated Results}\label{app:simulation}
We test the robustness of our probabilistic pair methodology by applying it to a mock dataset with the goal of recreating the observed weighted AGN fractions (Figure~\ref{fig:OBSC_ENH}) of our obscured AGN sample. We generate 250,000 galaxies in a ${\sim}2.4$\,deg$^2$ area, from which we draw the RA and DEC of each galaxy randomly. True redshifts, $z_\textrm{true}$, are drawn from a random uniform distribution between 0.5 and 3. To simulate increasing weighted AGN fractions at decreasing separation comparable to that of the observed obscured AGN enhancement, we then randomly assign AGN to galaxy pairs ($\Delta V < 1000$\,km\,s$^{-1}$) at 0.02, 0.01 and 0.005 probabilities if their projected separation is within 25\,kpc, 50\,kpc, or not (i.e., isolated), respectively.

We generate photometric redshift PDFs for each galaxy from a Gaussian distribution of mean $\mu_\textrm{sim}$ and standard deviation $\sigma_\textrm{sim}$. We fit Gaussian distributions of $\sigma$ equal to ${\sim}0.017$ and ${\sim}0.068$ (both $\mu \sim 0$) to the observed $\Delta z$ (i.e., $z_\textrm{spec} - z_\textrm{phot}$) and $\sigma_{P(z)}$ (i.e., half the upper minus lower PDF $68\%$ confidence intervals), respectively, from which we draw $\Delta z_i$ shifts and $\sigma_{P(z), i}$ such that $\mu_{\textrm{sim},i} = z_{\textrm{true},i} + \Delta z_i$ and $\sigma_{\textrm{sim},i} = \sigma_{P(z),i}$ for galaxy $i$.

Finally, we calculate the weighted AGN fractions from the simulated PDFs, which we compare to the true statistics of the mock sample (i.e., from $z_\textrm{true}$) as a function of projected separation in Figure~\ref{fig:sim_pairs}. Determining the AGN enhancement via PDF convolutions indeed recovers an increase in the weighted AGN fraction from ${\sim}0.5\%$ at projected separations less than 50\,kpc to ${\sim}10\%$ at projected separations less than 25\,kpc, just as in the obscured AGN sample (Figure~\ref{fig:OBSC_ENH}).

However, based on this simulated experiment, the convolution approach underestimates the true AGN fraction of close galaxy pairs (within 25\,kpc) in the mock dataset by ${\sim}40\%$. Given that the X-ray sample has a much higher spectroscopic redshift coverage (i.e., roughly one third have spectrosopic redshifts), and therefore is less affected by uncertain PDFs, and we observe no enhancement at all for this sample (Figure~\ref{fig:XAGN_IRAGN}), the observed lack of a trend with projected separation is expected to be robust. For the obscured AGN sample (Figure~\ref{fig:OBSC_ENH}), our simulations indicate that our finding may be moderately \textit{underestimating} the true AGN enhancements (albeit this is captured somewhat with that large uncertainties) as true and control pair samples inevitably become less clean with a higher fraction of photometric redshifts.

\begin{figure}
    \includegraphics[width=\columnwidth]{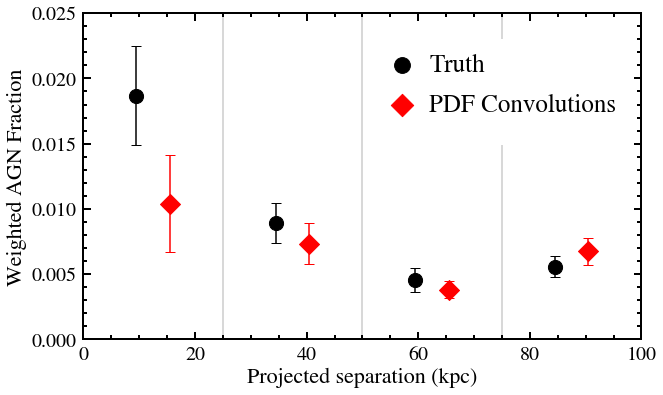}
    \caption{Weighted AGN fraction of simulated galaxy pairs binned by projected separation. The black circles correspond to the weighted AGN fraction ($\mathcal{P}_\textrm{pair}$ equal to 0 or 1) calculated from the true redshifts (i.e., $z_\textrm{true}$). The red diamonds correspond to the weighted AGN fraction determined using simulated PDFs and the pair convolution method detailed in this work. Standard errors for both groups are calculated from a common bootstrap technique \citep{Efron79, Efron81}.}
    \label{fig:sim_pairs}
\end{figure}

\section{Supplementary Information}
\label{app:tables}

This contains all tabulated data plotted in Section~\ref{sec:Results}. This includes the weighted AGN fractions and enhancements in bins of projected separation for X-ray AGN (Table~\ref{tab:all_XAGN}), that for X-ray AGN split by luminosity and redshift (Table~\ref{tab:LX_cuts}), IR AGN (Table~\ref{tab:all_IRAGN}) and obscured AGN (defined as those only selected in the IR; Table~\ref{tab:obsc_AGN_enh}). Obscuration fraction (ratio of obscured to X-ray AGN) as a function of projected separation is also tabulated in Table~\ref{tab:obsc_frac}.

\begin{landscape}
 \begin{table}
    \centering
    \caption{Enhancement of X-ray AGN ($L_X > 10^{42}$\,erg\,s$^{-1}$) in major (mass ratio up to 4:1) and minor galaxy pairs (mass ratios of 4:1 to 10:1) at $0.5 < z < 3$. N ($\mathcal{P}_\textrm{pair} > 0.01$) gives the number of unique galaxy pairs with pair probability ($\mathcal{P}_\textrm{pair}$; Equation~\ref{eq:trueP}) greater than 0.01. $\sum \mathcal{P}_\textrm{pair}$ and $\sum \mathcal{P}_\textrm{pair} \times N_\textrm{AGN}$ correspond to the sum of all pair probabilities and the sum of all AGN in pairs (0, 1, or 2) weighted by those pair probabilities, respectively. The weighted AGN fraction is the quotient of the previous two sums (Equation~\ref{eq:weightedAGNfrac}), and AGN enhancement is computed as the ratio of the weighted AGN fraction of true pairs to that of the control pairs (Equation~\ref{eq:AGN_enh}). The $z_\textrm{spec} + z_\textrm{phot}$ and $z_\textrm{phot}$ only sections correspond to results with and without secure spectroscopic redshifts ($\delta(z_\textrm{spec})$) in place of photometric redshift probability distribution functions when available. All tabulated values are plotted in the left panels of Figure~\ref{fig:XAGN_IRAGN}.}
    \label{tab:all_XAGN}
    \renewcommand{\arraystretch}{1.1} 
    \begin{tabular}{ c c l c c c c c c c c c }
    \hline \hline \\[-0.3cm]
        & ~ & ~ & \multicolumn{4}{c}{Major pairs} & ~ & \multicolumn{4}{c}{Minor pairs} \\[0.1cm] \cline{4-7} \cline{9-12} \\[-0.3cm]
        & ~ & ~ & $0 < r_p < 25$ & $25 < r_p < 50$ & $50 < r_p < 75$ & $75 < r_p < 100$ & ~ & $0 < r_p < 25$ & $25 < r_p < 50$ & $50 < r_p < 75$ & $75 < r_p < 100$  \\[0.1cm] \hline \\[-0.3cm]
        \multirow{9}{*}{\makecell[c]{$z_\textrm{spec}$ \\ + \\ $z_\textrm{phot}$}} & \multirow{4}{*}{\makecell[c]{True \\ pairs}} & N ($\mathcal{P}_\textrm{pair} > 0.01$) & 2599 & 5126 & 6482 & 8088 & ~ & 1133 & 2299 & 2824 & 3556  \\
         & ~ & $\sum \mathcal{P}_\textrm{pair}$ & 357.2 & 607.1 & 706.9 & 849.7 & ~ & 138.6 & 265.9 & 306.1 & 373.7  \\
         & ~ & $\sum \mathcal{P}_\textrm{pair} \times N_\textrm{AGN}$ & 26.0 & 42.6 & 43.2 & 60.2 & ~ & 9.1 & 20.0 & 23.9 & 29.9  \\
         & ~ & Weighted AGN fraction (\%) & $7.28\pm0.90$ & $7.01\pm0.72$ & $6.11\pm0.66$ & $7.09\pm0.65$ & ~ & $6.57\pm1.06$ & $7.51\pm1.08$ & $7.80\pm0.84$ & $8.01\pm0.85$  \\ \cline{2-12}
         & \multirow{4}{*}{\makecell[c]{Control \\ pairs}} & N ($\mathcal{P}_\textrm{pair} > 0.01$) & 7792 & 15343 & 19473 & 24270 & ~ & 3398 & 6896 & 8477 & 10671  \\
         & ~ & $\sum \mathcal{P}_\textrm{pair}$ & 1067.4 & 1813.8 & 2117.2 & 2543.5 & ~ & 413.5 & 795.8 & 916.8 & 1117.5  \\
         & ~ & $\sum \mathcal{P}_\textrm{pair} \times N_\textrm{AGN}$ & 75.9 & 110.0 & 116.1 & 151.9 & ~ & 27.7 & 61.1 & 66.4 & 80.2  \\
         & ~ & Weighted AGN fraction (\%) & $7.12\pm0.52$ & $6.06\pm0.36$ & $5.48\pm0.32$ & $5.97\pm0.29$ & ~ & $6.70\pm0.70$ & $7.68\pm0.55$ & $7.24\pm0.54$ & $7.18\pm0.42$  \\ \cline{2-12}
         & ~ & AGN enhancement & $1.02\pm0.15$ & $1.16\pm0.14$ & $1.12\pm0.14$ & $1.19\pm0.12$ & ~ & $0.98\pm0.19$ & $0.98\pm0.16$ & $1.08\pm0.14$ & $1.12\pm0.14$  \\ \hline
        \multirow{9}{*}{\makecell[c]{$z_\textrm{phot}$ \\ only}} & \multirow{4}{*}{\makecell[c]{True \\ pairs}} & N ($\mathcal{P}_\textrm{pair} > 0.01$) & 2639 & 5226 & 6652 & 8310 & ~ & 1142 & 2314 & 2923 & 3646  \\
         & ~ & $\sum \mathcal{P}_\textrm{pair}$ & 302.7 & 550.5 & 654.6 & 770.5 & ~ & 120.9 & 230.9 & 269.6 & 338.4  \\
         & ~ & $\sum \mathcal{P}_\textrm{pair} \times N_\textrm{AGN}$ & 17.9 & 29.2 & 33.2 & 37.0 & ~ & 8.6 & 15.9 & 17.9 & 23.2  \\
         & ~ & Weighted AGN fraction (\%) & $5.92\pm0.55$ & $5.30\pm0.40$ & $5.07\pm0.35$ & $4.81\pm0.34$ & ~ & $7.09\pm0.92$ & $6.89\pm0.83$ & $6.64\pm0.64$ & $6.84\pm0.59$  \\ \cline{2-12}
         & \multirow{4}{*}{\makecell[c]{Control \\ pairs}} & N ($\mathcal{P}_\textrm{pair} > 0.01$) & 7902 & 15637 & 19987 & 24936 & ~ & 3428 & 6941 & 8778 & 10935  \\
         & ~ & $\sum \mathcal{P}_\textrm{pair}$ & 901.9 & 1645.8 & 1958.1 & 2306.2 & ~ & 361.5 & 689.9 & 805.8 & 1013.2  \\
         & ~ & $\sum \mathcal{P}_\textrm{pair} \times N_\textrm{AGN}$ & 57.7 & 84.2 & 100.7 & 114.8 & ~ & 25.8 & 53.7 & 55.4 & 71.9  \\
         & ~ & Weighted AGN fraction (\%) & $6.40\pm0.39$ & $5.11\pm0.27$ & $5.14\pm0.25$ & $4.98\pm0.21$ & ~ & $7.15\pm0.58$ & $7.78\pm0.48$ & $6.88\pm0.39$ & $7.10\pm0.33$  \\ \cline{2-12}
         & ~ & AGN enhancement & $0.93\pm0.10$ & $1.04\pm0.09$ & $0.99\pm0.08$ & $0.97\pm0.08$ & ~ & $0.99\pm0.15$ & $0.89\pm0.12$ & $0.96\pm0.11$ & $0.96\pm0.09$ \\ \hline
    \end{tabular}
 \end{table}
\end{landscape}

\begin{landscape}
 \begin{table}
    \centering
    \caption{Enhancement of IR AGN \citep{Donley12} in major (mass ratio up to 4:1) and minor galaxy pairs (mass ratios of 4:1 to 10:1) at $0.5 < z < 3$. Row definitions follow that of Table~\ref{tab:all_XAGN}, and all tabulated values are plotted in the right panels of Figure~\ref{fig:XAGN_IRAGN}.}
    \label{tab:all_IRAGN}
    \renewcommand{\arraystretch}{1.1} 
    \begin{tabular}{ c c l c c c c c c c c c }
    \hline \hline \\[-0.3cm]
        & ~ & ~ & \multicolumn{4}{c}{Major pairs} & ~ & \multicolumn{4}{c}{Minor pairs} \\[0.1cm] \cline{4-7} \cline{9-12} \\[-0.3cm]
        & ~ & ~ & $0 < r_p < 25$ & $25 < r_p < 50$ & $50 < r_p < 75$ & $75 < r_p < 100$ & ~ & $0 < r_p < 25$ & $25 < r_p < 50$ & $50 < r_p < 75$ & $75 < r_p < 100$  \\[0.1cm] \hline \\[-0.3cm]
         \multirow{9}{*}{\makecell[c]{$z_\textrm{spec}$ \\ + \\ $z_\textrm{phot}$}} & \multirow{4}{*}{\makecell[c]{True \\ pairs}} & N ($\mathcal{P}_\textrm{pair} > 0.01$) & 2599 & 5126 & 6482 & 8088 & ~ & 1133 & 2299 & 2824 & 3556  \\
        & ~ & $\sum \mathcal{P}_\textrm{pair}$ & 357.2 & 607.1 & 706.9 & 849.7 & ~ & 138.6 & 265.9 & 306.1 & 373.7  \\
        & ~ & $\sum \mathcal{P}_\textrm{pair} \times N_\textrm{AGN}$ & 7.6 & 11.8 & 12.4 & 8.5 & ~ & 2.3 & 5.8 & 5.8 & 7.2  \\
        & ~ & Weighted AGN fraction (\%) & $2.12\pm0.42$ & $1.95\pm0.36$ & $1.76\pm0.33$ & $1.00\pm0.18$ & ~ & $1.67\pm0.41$ & $2.18\pm0.61$ & $1.88\pm0.37$ & $1.94\pm0.41$  \\ \cline{2-12}
        & \multirow{4}{*}{\makecell[c]{Control \\ pairs}} & N ($\mathcal{P}_\textrm{pair} > 0.01$) & 7792 & 15343 & 19473 & 24270 & ~ & 3398 & 6896 & 8477 & 10671  \\
        & ~ & $\sum \mathcal{P}_\textrm{pair}$ & 1067.4 & 1813.8 & 2117.2 & 2543.5 & ~ & 413.5 & 795.8 & 916.8 & 1117.5  \\
        & ~ & $\sum \mathcal{P}_\textrm{pair} \times N_\textrm{AGN}$ & 17.2 & 25.5 & 25.4 & 31.0 & ~ & 7.7 & 11.4 & 16.2 & 18.2  \\
        & ~ & Weighted AGN fraction (\%) & $1.61\pm0.24$ & $1.40\pm0.14$ & $1.20\pm0.11$ & $1.22\pm0.11$ & ~ & $1.85\pm0.31$ & $1.44\pm0.25$ & $1.77\pm0.20$ & $1.62\pm0.17$  \\ \cline{2-12}
        & ~ & AGN enhancement & $1.32\pm0.33$ & $1.39\pm0.29$ & $1.47\pm0.31$ & $0.82\pm0.17$ & ~ & $0.90\pm0.27$ & $1.51\pm0.50$ & $1.06\pm0.24$ & $1.19\pm0.28$  \\ \hline
        \multirow{9}{*}{\makecell[c]{$z_\textrm{phot}$ \\ only}} & \multirow{4}{*}{\makecell[c]{True \\ pairs}} & N ($\mathcal{P}_\textrm{pair} > 0.01$) & 2639 & 5226 & 6652 & 8310 & ~ & 1142 & 2314 & 2923 & 3646  \\
        & ~ & $\sum \mathcal{P}_\textrm{pair}$ & 302.7 & 550.5 & 654.6 & 770.5 & ~ & 120.9 & 230.9 & 269.6 & 338.4  \\
        & ~ & $\sum \mathcal{P}_\textrm{pair} \times N_\textrm{AGN}$ & 6.4 & 7.5 & 8.5 & 6.6 & ~ & 2.2 & 3.3 & 5.2 & 6.5  \\
        & ~ & Weighted AGN fraction (\%) & $2.10\pm0.33$ & $1.37\pm0.19$ & $1.29\pm0.19$ & $0.86\pm0.13$ & ~ & $1.80\pm0.47$ & $1.41\pm0.29$ & $1.91\pm0.31$ & $1.91\pm0.30$  \\ \cline{2-12}
        & \multirow{4}{*}{\makecell[c]{Control \\ pairs}} & N ($\mathcal{P}_\textrm{pair} > 0.01$) & 7902 & 15637 & 19987 & 24936 & ~ & 3428 & 6941 & 8778 & 10935  \\
        & ~ & $\sum \mathcal{P}_\textrm{pair}$ & 901.9 & 1645.8 & 1958.1 & 2306.2 & ~ & 361.5 & 689.9 & 805.8 & 1013.2  \\
        & ~ & $\sum \mathcal{P}_\textrm{pair} \times N_\textrm{AGN}$ & 16.6 & 23.3 & 26.8 & 29.5 & ~ & 7.7 & 12.6 & 15.2 & 22.3  \\
        & ~ & Weighted AGN fraction (\%) & $1.84\pm0.23$ & $1.42\pm0.15$ & $1.37\pm0.12$ & $1.28\pm0.10$ & ~ & $2.13\pm0.31$ & $1.82\pm0.23$ & $1.89\pm0.19$ & $2.20\pm0.19$  \\ \cline{2-12}
        & ~ & AGN enhancement & $1.14\pm0.23$ & $0.97\pm0.17$ & $0.95\pm0.16$ & $0.67\pm0.11$ & ~ & $0.84\pm0.25$ & $0.78\pm0.19$ & $1.01\pm0.19$ & $0.87\pm0.16$ \\ \hline
    \end{tabular}
 \end{table}
\end{landscape}

\begin{landscape}
 \begin{table}
    \centering
    \caption{Enhancement of moderate $L_X$ ($L_X = 10^{43.2{-}43.7}$\,erg\,s$^{-1}$) and high $L_X$ AGN ($L_X > 10^{43.7}$\,erg\,s$^{-1}$; see Figure~\ref{fig:LX_sample}) in major galaxy pairs (mass ratio up to 4:1) as a function of projected separation, $r_p$, using available spectroscopic redshifts. The table is split by redshift such that the first section corresponds to the full complete redshift range of each sample ($0.5 < z < 2$ and $0.5 < z < 3$ for moderate $L_X$ and high $L_X$ AGN, respectively), the second to $1 < z < 2$ and the third to $2 < z < 3$, which only the high $L_X$ AGN sample populates. Row definitions follow that of Table~\ref{tab:all_XAGN}, and all tabulated values are plotted in Figure~\ref{fig:MIDLX_HILX}.}
    \label{tab:LX_cuts}
    \renewcommand{\arraystretch}{1.1} 
    \begin{tabular}{ c c l c c c c c c c c c }
    \hline \hline \\[-0.3cm]
       & ~ & ~ & \multicolumn{4}{c}{Moderate $L_X$} & ~ & \multicolumn{4}{c}{High $L_X$} \\[0.1cm] \cline{4-7} \cline{9-12} \\[-0.3cm]
        & ~ & ~ & $0 < r_p < 25$ & $25 < r_p < 50$ & $50 < r_p < 75$ & $75 < r_p < 100$ & ~ & $0 < r_p < 25$ & $25 < r_p < 50$ & $50 < r_p < 75$ & $75 < r_p < 100$  \\[0.1cm] \hline \\[-0.3cm]
        \multirow{9}{*}{\makecell[c]{Complete \\ redshfit \\ range}} & \multirow{4}{*}{\makecell[c]{True \\ pairs}} & N ($\mathcal{P}_\textrm{pair} > 0.01$) & 1734 & 3611 & 4549 & 5642 & ~ & 2599 & 5126 & 6482 & 8088  \\
        & ~ & $\sum \mathcal{P}_\textrm{pair}$ & 256.8 & 459.2 & 534.4 & 643.4 & ~ & 357.2 & 607.1 & 706.9 & 849.7  \\
        & ~ & $\sum \mathcal{P}_\textrm{pair} \times N_\textrm{AGN}$ & 4.7 & 11.0 & 11.6 & 13.7 & ~ & 6.2 & 7.3 & 9.2 & 10.8  \\
        & ~ & Weighted AGN fraction (\%) & $1.82\pm0.60$ & $2.39\pm0.40$ & $2.16\pm0.48$ & $2.13\pm0.38$ & ~ & $1.73\pm0.35$ & $1.20\pm0.25$ & $1.30\pm0.25$ & $1.27\pm0.19$  \\ \cline{2-12}
        & \multirow{4}{*}{\makecell[c]{Control \\ pairs}} & N ($\mathcal{P}_\textrm{pair} > 0.01$) & 5193 & 10813 & 13671 & 16923 & ~ & 7792 & 15343 & 19473 & 24270  \\
        & ~ & $\sum \mathcal{P}_\textrm{pair}$ & 767.6 & 1371.6 & 1600.3 & 1926.4 & ~ & 1067.4 & 1813.8 & 2117.2 & 2543.5  \\
        & ~ & $\sum \mathcal{P}_\textrm{pair} \times N_\textrm{AGN}$ & 15.8 & 30.1 & 29.0 & 34.2 & ~ & 15.0 & 26.0 & 26.7 & 30.5  \\
        & ~ & Weighted AGN fraction (\%) & $2.06\pm0.32$ & $2.20\pm0.25$ & $1.81\pm0.19$ & $1.78\pm0.17$ & ~ & $1.41\pm0.20$ & $1.44\pm0.15$ & $1.26\pm0.12$ & $1.20\pm0.11$  \\ \cline{2-12}
        & ~ & AGN enhancement & $0.88\pm0.32$ & $1.09\pm0.22$ & $1.19\pm0.29$ & $1.20\pm0.25$ & ~ & $1.23\pm0.30$ & $0.83\pm0.20$ & $1.03\pm0.22$ & $1.06\pm0.19$  \\ \hline
        \multirow{9}{*}{\makecell[c]{$1 < z < 2$}} & \multirow{4}{*}{\makecell[c]{True \\ pairs}} & N ($\mathcal{P}_\textrm{pair} > 0.01$) & 1238 & 2548 & 3216 & 4012 & ~ & 1238 & 2548 & 3216 & 4012  \\
        & ~ & $\sum \mathcal{P}_\textrm{pair}$ & 147.7 & 270.9 & 315.2 & 382.5 & ~ & 147.7 & 270.9 & 315.2 & 382.5  \\
        & ~ & $\sum \mathcal{P}_\textrm{pair} \times N_\textrm{AGN}$ & 3.3 & 8.8 & 9.1 & 9.7 & ~ & 2.9 & 4.0 & 5.4 & 5.4  \\
        & ~ & Weighted AGN fraction (\%) & $2.23\pm0.80$ & $3.23\pm0.51$ & $2.89\pm0.68$ & $2.54\pm0.50$ & ~ & $1.98\pm0.56$ & $1.48\pm0.43$ & $1.72\pm0.52$ & $1.41\pm0.31$  \\ \cline{2-12}
        & \multirow{4}{*}{\makecell[c]{Control \\ pairs}} & N ($\mathcal{P}_\textrm{pair} > 0.01$) & 3704 & 7618 & 9673 & 12025 & ~ & 3704 & 7618 & 9673 & 12025  \\
        & ~ & $\sum \mathcal{P}_\textrm{pair}$ & 441.1 & 809.9 & 944.3 & 1145.0 & ~ & 441.1 & 809.9 & 944.3 & 1145.0  \\
        & ~ & $\sum \mathcal{P}_\textrm{pair} \times N_\textrm{AGN}$ & 11.9 & 22.5 & 22.2 & 25.9 & ~ & 6.4 & 12.9 & 13.9 & 15.8  \\
        & ~ & Weighted AGN fraction (\%) & $2.69\pm0.41$ & $2.78\pm0.34$ & $2.36\pm0.27$ & $2.26\pm0.21$ & ~ & $1.44\pm0.31$ & $1.59\pm0.26$ & $1.47\pm0.17$ & $1.38\pm0.18$  \\ \cline{2-12}
        & ~ & AGN enhancement & $0.83\pm0.32$ & $1.16\pm0.23$ & $1.23\pm0.32$ & $1.12\pm0.25$ & ~ & $1.37\pm0.49$ & $0.93\pm0.31$ & $1.17\pm0.38$ & $1.02\pm0.26$ \\ \hline
        \multirow{9}{*}{\makecell[c]{$2 < z < 3$}} & \multirow{4}{*}{\makecell[c]{True \\ pairs}} & N ($\mathcal{P}_\textrm{pair} > 0.01$) & ~ & ~ & ~ & ~ & ~ & 855 & 1509 & 1920 & 2439  \\
        & ~ & $\sum \mathcal{P}_\textrm{pair}$ & ~ & ~ & ~ & ~ & ~ & 100.1 & 147.4 & 171.1 & 205.9  \\
        & ~ & $\sum \mathcal{P}_\textrm{pair} \times N_\textrm{AGN}$ & ~ & ~ & ~ & ~ & ~ & 3.122 & 3.247 & 3.268 & 5.376  \\
        & ~ & Weighted AGN fraction (\%) & ~ & ~ & ~ & ~ & ~ & $3.12\pm0.98$ & $2.20\pm0.74$ & $1.91\pm0.36$ & $2.61\pm0.56$  \\ \cline{2-12}
        & \multirow{4}{*}{\makecell[c]{Control \\ pairs}} & N ($\mathcal{P}_\textrm{pair} > 0.01$) & ~ & ~ & ~ & ~ & ~ & 2570 & 4512 & 5763 & 7326  \\
        & ~ & $\sum \mathcal{P}_\textrm{pair}$ & ~ & ~ & ~ & ~ & ~ & 298.6 & 440.8 & 512.7 & 615.7  \\
        & ~ & $\sum \mathcal{P}_\textrm{pair} \times N_\textrm{AGN}$ & ~ & ~ & ~ & ~ & ~ & 7.4 & 9.9 & 8.5 & 11.2  \\
        & ~ & Weighted AGN fraction (\%) & ~ & ~ & ~ & ~ & ~ & $2.49\pm0.34$ & $2.25\pm0.33$ & $1.65\pm0.23$ & $1.82\pm0.27$  \\ \cline{2-12}
        & ~ & AGN enhancement & ~ & ~ & ~ & ~ & ~ & $1.25\pm0.43$ & $0.98\pm0.36$ & $1.16\pm0.27$ & $1.44\pm0.37$ \\ \hline
    \end{tabular}
 \end{table}
\end{landscape}

\begin{table*}
    \centering
    \caption{Obscuration fraction (ratio of obscured AGN to all X-ray-selected AGN; Equation~\ref{eq:obsc_frac}) of major galaxy pairs (mass ratio up to 4:1) as a function of projected separation, $r_p$, using available spectroscopic redshifts. Row definitions follow the scheme outlined in Table~\ref{tab:all_XAGN}. The sums of weighted obscured and X-ray AGN are reported separately ($\sum \mathcal{P}_\textrm{pair} \times N_\textrm{IR-only AGN}$ and $\sum \mathcal{P}_\textrm{pair} \times N_\textrm{X-ray AGN}$), and obscuration fraction is computed as the ratio of those sums. All tabulated values are plotted in Figure~\ref{fig:OBSC_FRAC2}.}
    \label{tab:obsc_frac}
    \renewcommand{\arraystretch}{1.1} 
    \begin{tabular}{ c l c c c c }
    \hline \hline \\[-0.3cm]
        & ~ & $0 < r_p < 25$ & $25 < r_p < 50$ & $50 < r_p < 75$ & $75 < r_p < 100$ \\[0.1cm] \hline \\[-0.3cm]
        \multirow{4}{*}{\makecell[c]{True \\ pairs}} & N ($\mathcal{P}_\textrm{pair} > 0.01$) & 2599 & 5126 & 6482 & 8088  \\
        & $\sum \mathcal{P}_\textrm{pair}$ & 357.2 & 607.1 & 706.9 & 849.7  \\
        & $\sum \mathcal{P}_\textrm{pair} \times N_\textrm{IR-only AGN}$ & 3.8 & 4.8 & 4.8 & 4.3  \\
        & $\sum \mathcal{P}_\textrm{pair} \times N_\textrm{X-ray AGN}$ & 26.0 & 42.6 & 43.2 & 60.2  \\
        & Obscuration fraction (\%) & $14.63\pm3.83$ & $11.23\pm2.31$ & $11.11\pm2.56$ & $7.17\pm1.63$  \\ \hline
        \multirow{4}{*}{\makecell[c]{Control \\ pairs}} & N ($\mathcal{P}_\textrm{pair} > 0.01$) & 7792 & 15343 & 19473 & 24270  \\
        & $\sum \mathcal{P}_\textrm{pair}$ & 1067.4 & 1813.8 & 2117.2 & 2543.5  \\
        & $\sum \mathcal{P}_\textrm{pair} \times N_\textrm{IR-only AGN}$ & 5.4 & 9.2 & 10.9 & 13.3  \\
        & $\sum \mathcal{P}_\textrm{pair} \times N_\textrm{X-ray AGN}$ & 75.9 & 110.0 & 116.1 & 151.9  \\
        & Obscuration fraction (\%) & $7.17\pm1.39$ & $8.39\pm1.29$ & $9.42\pm1.21$ & $8.74\pm1.40$ \\ \hline
    \end{tabular}
\end{table*}

\begin{table*}
    \centering
    \caption{Enhancement of obscured AGN (selected exclusively in the IR; see Section~\ref{sec:Obscured_AGN}) in major galaxy pairs (mass ratio up to 4:1) as a function of projected separation, $r_p$, using available spectroscopic redshifts. Row definitions follow that of Table~\ref{tab:all_XAGN}, and all tabulated values are plotted in Figure~\ref{fig:OBSC_ENH}.}
    \label{tab:obsc_AGN_enh}
    \renewcommand{\arraystretch}{1.1} 
    \begin{tabular}{ c l c c c c }
    \hline \hline \\[-0.3cm]
        & ~ & $0 < r_p < 25$ & $25 < r_p < 50$ & $50 < r_p < 75$ & $75 < r_p < 100$ \\[0.1cm] \hline \\[-0.3cm]
        \multirow{4}{*}{\makecell[c]{True \\ pairs}} & N ($\mathcal{P}_\textrm{pair} > 0.01$) & 2599 & 5126 & 6482 & 8088  \\
        & $\sum \mathcal{P}_\textrm{pair}$ & 357.2 & 607.1 & 706.9 & 849.7  \\
        & $\sum \mathcal{P}_\textrm{pair} \times N_\textrm{AGN}$ & 3.8 & 4.8 & 4.8 & 4.3  \\
        & Weighted AGN fraction (\%) & $1.06\pm0.25$ & $0.79\pm0.15$ & $0.68\pm0.15$ & $0.51\pm0.09$ \\ \hline
        \multirow{4}{*}{\makecell[c]{Control \\ pairs}} & N ($\mathcal{P}_\textrm{pair} > 0.01$) & 7792 & 15343 & 19473 & 24270  \\
        & $\sum \mathcal{P}_\textrm{pair}$ & 1067.4 & 1813.8 & 2117.2 & 2543.5  \\
        & $\sum \mathcal{P}_\textrm{pair} \times N_\textrm{AGN}$ & 5.4 & 9.2 & 10.9 & 13.3  \\
        & Weighted AGN fraction (\%) & $0.51\pm0.09$ & $0.51\pm0.07$ & $0.52\pm0.06$ & $0.52\pm0.07$ \\ \hline
        & AGN enhancement & $2.08\pm0.61$ & $1.55\pm0.36$ & $1.31\pm0.32$ & $0.97\pm0.22$ \\ \hline
    \end{tabular}
\end{table*}

\bsp	
\label{lastpage}
\end{document}